\documentclass[12pt]{article}
\usepackage{amsmath,amssymb,cite}
\usepackage{graphicx}
\usepackage{verbatim}
\usepackage{amsthm}
\usepackage{ifpdf}
\makeatletter

    \@addtoreset{equation}{section}
\makeatother
\newcommand{\One}{{\hbox{{\rm 1{\hbox to 1.5pt{\hss\rm1}}}}}}

\newcommand{\vev}[1]{\langle#1\rangle}
\newcommand{\CA}{{\cal A}}

\newcommand{\CF}{{\cal F}}

\newcommand{\CN}{{\cal N}}

\newcommand{\CL}{{\cal L}}
\newcommand{\CO}{{\cal O}}
\newcommand{\CP}{{\cal P}}

\newcommand{\td}{\tilde{D}}
\newcommand{\la}{\lambda}

\newcommand\be{\begin{equation}}
\newcommand\ee{\end{equation}}
\newcommand\bea{\begin{eqnarray}}
\newcommand\eea{\end{eqnarray}}
\newcommand\ba{\(\begin{array}}
\newcommand\ea{\end{array})\ }
\newcommand\nn{\nonumber}

\begin{document}
\thispagestyle{empty}
\vspace*{2cm}
\begin{center}
 {\LARGE {Parametric dependence of irregular conformal block}}
\vskip2cm
{\large 
{Sang-Kwan Choi\footnote{email: hermit1231@gmail.com}
}
and\hspace{2mm} Chaiho Rim\footnote{email: rimpine@sogang.ac.kr}
}
\vskip.5cm
{\it Department of Physics and Center for Quantum Spacetime, \\
Sogang University, Seoul 121-742 Korea}
\end{center}
%
%
 \vskip2cm
\begin{abstract}
Irregular conformal block is 
an important tool to study a new type of conformal theories,
which can be constructed as the colliding limit of the regular conformal block.
The irregular conformal block is realized as
the  $\beta$-deformed Penner matrix model whose partition function 
is regarded as the inner product of the irregular modules.
The parameter dependence of the inner product is obtained 
explicitly 
using the loop equation 
with close attention to singularities in the parameter space.
It is noted that the exact singular structure of the parameter space
in general can be found using a very simple and powerful method 
which uses the flow equations of the partition function
together with the hierarchical structure of the singularity. 
This method  gives the exact expression  to all orders of large $N$ expansion
without using the explicit contour integral of the filling fraction.

\end{abstract}
\newpage
\tableofcontents


\section{Introduction}
In conformal field theory one constructs the  lowest weight representation
using the Virasoro lowest  weight state $|\Delta \rangle$ 
\cite{cft}. 
Recently, however, so called irregular vector 
was introduced in \cite{G2009} in connection with 
the asymptotically free  $\CN=2$ gauge theories.   
The irregular vector  $ |I^{(m)} \rangle$  is 
annihilated by positive Virasoro generators $L_\ell |I^{(m)} \rangle = 0 $ 
when $\ell> 2m>0 $,
but is  a simultaneous eigenstate of a  set of Virasoro generators 
\be
L_k |I^{(m)} \rangle = \Lambda_k |I^{(m)} \rangle \, ,\quad m \leq k \leq 2m 
\label{prop1}
\ee 
where $\Lambda_k$ is a non-vanishing eigenvalue. 
The existence of the irregular vector is consistent with the Virasoro algebra 
 $ \left[L_k, L_n\right]=(k-n) L_{k+n} $  with $k,n>0$.  

The irregular vector can be constructed 
as the superposition of lowest state together with its descendents.  
Explicit construction for small $m$ can be found in  
\cite{G2009, MMM2009, FJK2011}. 
A more detailed investigation 
is done in \cite{BMT2011,GT2012} in connection 
with Argyres-Douglas type gauge theories
\cite{AD1995,APSW1996}.  
It is also  noted in \cite{GT2012} that  the  irregular vector 
can be constructed in  the process of colliding limit of primary fields.
This is because  the colliding  limit of primary fields 
induces higher than 2 degree of  singularities 
in the operator product  expansion with the  energy-momentum tensor.
The higher singularity is the manifestation of the irregular vector. 

Let us construct a multi-point state  
$|R_m \rangle = \lim_{z_0 \to 0} \prod_{r=0}^{m} \Psi_{\Delta_r} (z_r) | 0 \rangle$
where $\Psi_{\Delta_r}(z_r)$ is the primary field with conformal dimension $\Delta_r$. 
We put $\Psi_{\Delta_0}(z_0)$ at the origin so that 
the lowest weight state $ |\Delta_0 \rangle = \lim_{z \to 0} \Psi_{\Delta_0}(z) |0 \rangle $
is obtained.  
The operator product expansion with the energy-momentum tensor $T(z)$ has the form
\be
T(z) |R_m \rangle = \sum_{r=0}^{m} \biggl( \frac{\Delta_r}{(z-z_r)^{2}}
+ \frac{1}{(z-z_r)}\frac{\partial}{\partial z_r}+\mathrm{regular\, \, terms} \biggr) |R_m \rangle\,.
\ee
The colliding limit is obtained if $z_r \to 0$ but $\Delta_r \to \infty$ so that 
there appear non-vanishing parameters; 
\be c_0=\sum_{r=0}^{m} \alpha_r \,,~~~
c_k= \sum_{r=0}^{m} \alpha_r \sum_{\substack{{0\leq s_1 < \cdots <s_k \leq m} \\{s_i \neq r}}}
\prod_{i=1}^{k} (-z_{s_i}) ~~{\rm when}~0 \leq k \leq m 
\label{c_k}
\ee  
where $\Delta_r = \alpha_r (Q-\alpha_r)$ is used.  
Then, higher singularities in the OPE with $T(z)$ are induced at the colliding limit  
\begin{align}
& T(z) |I^{(m)} \rangle = \biggl( \sum_{k=0}^{2m} \frac{\mathcal{L}_k}{z^{k+2}}
 +\frac{L_{-1}}{z}+\mathrm{reg.} \biggr) |I^{(m)} \rangle\,,
\label{OPE_irreg} \\
& |I^{(m)} \rangle = \lim_{z_r \to 0, \alpha_r \to \infty}  \prod_{0 \leq r < s \leq m} 
(z_r -z_s)^{2 \alpha_r \alpha_s} |R_m \rangle
\end{align}
where we compensate the position singularity of  $|R_m \rangle $ by applying 
the products of  $(z_r-z_s)$'s so that 
the irregular vector $|I^{(m)} \rangle $ is well-defined.  
The  induced operator $\mathcal{L}_k$  is the realization of the Virasoro algebra  on the 
irregular vector space.
\be
\mathcal{L}_k = \Lambda_k + \sum_{\ell \in \mathbb{N}} (\ell-k) c_{\ell}
\frac{\partial}{\partial c_{\ell-k}} \,,~~~
\Lambda_k=(k+1)Qc_k - \sum_{\ell=0}^{k} c_{\ell} c_{k-\ell}\,, 
\ee  
where the notation is used: $c_{\ell} \equiv 0$ unless $0 \leq k \leq m$. 
It is obvious that 
$\CL_k= \Lambda_k$ for $m \leq k \leq 2m$
and is consistent with \eqref{prop1}.

\begin{figure}
\centering
\includegraphics[width=0.5\textwidth]{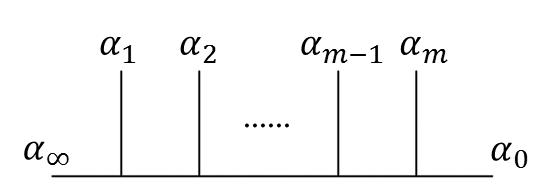}
\caption{Diagram of a $(m+2)$-point conformal block}
\label{f:reg}
\end{figure}

Our task is to investigate the parametric dependence of the irregular vector.
Instead of constructing directly as in \cite{G2009, MMM2009,FJK2011,GT2012},
we will use the Penner-type matrix model.
The Penner model was first introduced to find the Euler characteristic 
of moduli space of Riemann surfaces  with genus and punctures \cite{P1988}.  
Soon after, the Penner model turns out to be very useful to understand 
$c=1$ string theory \cite{DV1991, CDL1991} and is
further generalized \cite{DV2009} 
to obtain the conformal block of the Liouville theory 
inspired by AGT relation \cite{AGT}.
Note that $(m+2)$  correlation of vertex operators 
$ \left \langle \prod_{r=0}^{m+1} e^{2 \alpha_r \phi(z_r)} \right \rangle$ 
with the Liouville momentum $\alpha_a$ 
is evaluated perturbatively by expanding  the Liouville potential $e^{2 b \phi}$, 
$ \Big\langle \Big( \prod_{i=1}^{N} \int d\lambda_i d\bar{\lambda}_i e^{2b\phi (\lambda_i)}
\Big)  \prod_{r=0}^{m+1} e^{2 \alpha_r \phi(z_r)}
\Big\rangle_{\! 0} $
using the free correlation 
$\langle e^{2\alpha_1 \phi(z)} e^{2 \alpha_2 \phi(w)} \rangle_{\! 0} = |z-w|^{-4 \alpha_1 \alpha_2}$ 
with the neutrality condition  $ \sum_{a=0}^{m+1}  \alpha_a +  bN  =Q$. 
Here, $Q(=b+{1}/{b}) $ is the Liouville background charge. 
Then, the  conformal block $\CF_{m+2} (\left\{ e^{\alpha_k \phi(z_k)} \right\} )$  (figure \ref{f:reg})  
is identified with 
$Z_N  \times \prod_{0\le a <b \le m+1} (z_a -z_b)^{-2 \alpha_a \alpha_b} $ 
which defines the $\beta$-deformed Penner-type matrix model :
\be
Z_N  = \int \prod_{i=1}^N d\lambda_i \Delta(\la)^ {2 \beta} 
e^{ -\frac{\sqrt{\beta}}g \sum_i V(\lambda_i)} 
\label{beta-penner}
\ee
where  $\Delta(\la)=\prod_{i<j} (\la_i-\la_j)$ 
is the Vandermonde determinant.
The potential is given as  the sum of logarithmic terms: 
$ V(z) = -\sum_{a=0}^{m}   \alpha_a   \log (z -z_a ) $. 
To make  large $N$ expansion possible, we rescale $\alpha_a \to \alpha_a / \hbar  $  and 
rename $\beta \equiv -b^2$ and $\hbar \equiv -2 i g$ 
assuming  $\hbar =\CO(1/N)$ and $\alpha_a = \CO(1)$.
This $\beta$-deformed Penner type matrix is proved very useful to reproduce 
the Nekrasov partition function \cite{EM2009, NR2011}.

\begin{figure}
\centering
\includegraphics[width=0.5\textwidth]{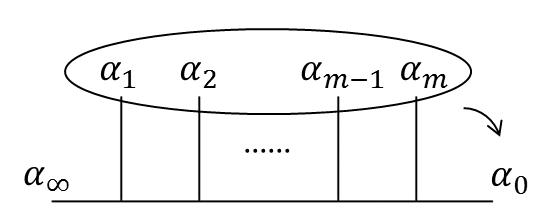}
\caption{Schematic diagram of $\vev{\Delta|I^{(m)}}$.
The irregular singularity with rank $m$ is developed 
when $m$-intermediate points are taken to one point at origin in a correlated way.}
\label{f:irreg}
\end{figure}

If we put $z_{m+1}=\infty$, we may view the conformal block
 $\CF_{m+2} (\left\{ e^{\alpha_k \phi(z_k)} \right\} )$ 
 as the inner product   $\vev{\Delta_\infty |R_m}$.
The out-state $\langle \Delta_\infty|$ has the Liouville momentum $\alpha_\infty$
and is the hermitian conjugation of the in-state $|\Delta_0 \rangle$ 
with $\alpha_0$. 
At the colliding limit (figure \ref{f:irreg}),  $|R_m \rangle$ 
becomes the irregular vector. 
With the compensating product of  $(z_a-z_b)$'s 
in  $\CF_{m+2} (\left\{ e^{\alpha_k \phi(z_k)} \right\} )$,  
$Z_N$  becomes  $\vev{\Delta|I^{(m)}} $
and the potential is given as the sum of logarithmic and 
inverse power terms
\be
V(z)= -c_0 \log z+  \sum_{k=1}^{m} \frac{c_k}{k z^k}\,.
\label{potential-m}
\ee 
We will use the notation $Z_N^{(m)}$  for 
the partition function with this  potential \eqref{potential-m} 
to distinguish from the one \eqref{beta-penner} 
which has logarithmic potentials only. 

Suppose one changes the integral variables $\lambda_i \rightarrow 1/\la_i$
in the partition function $Z_N^{(m)}$. 
Under the change of variable, 
the measure changes as $ \prod_{i=1}^N d \lambda_i  
\to  \prod_{i=1}^N {d \la_i}/{\la_i^2}$ and the Vandermonde determinant  
as $ \Delta(\lambda) \to 
\Delta(\la)   \prod_{k=1}^N \la_k^{ (1-N)}$. 
Exponentiating these extra factors into the potential term 
and using the neutrality condition 
one has the partition function with a slightly different potential 
\begin{equation}
V(z)=-\alpha_{\infty} \log z+\sum_{k=1}^{m} \frac{c_k z^k}{k} \,.
\end{equation}
We may interpret the resulting partition function as the hermitian conjugation 
satisfying the relation 
$\vev{ \Delta(\alpha_\infty) |I^{(m)} (\{c_k \})  }
=\vev{I^{(m)}(\{c_k\}) | \Delta  (\alpha_\infty)}$ \cite{NR2012}.

This paper is organized as follows. 
In section \ref{sec:solve} we evaluate the Penner-type partition function 
at the colliding limit.  
The evaluation is done using  the loop equation and the flow equations 
which was successively used in \cite{NR2012}.  
Here the filling fraction is used to replace all the unknown quantities.
In this way one finds the inner product between irregular vectors  
as well as the product between regular and  irregular vectors. 

In section \ref{sec:hier}, we present a simpler way to find the singular structure of 
the partition function, the singularity in the sense of the parameter space. 
This is achieved using the fact that the inner product can be put in a  hierarchical form.
The unknowns in the  flow equations are written in terms of power series of the 
properly defined parameters.
The self-consistency condition for the  flow equations provide
a certain set of recursion relations. 
Its initial data is  trivially given from the loop equation.  
The merit of this approach is that one does not have 
to evaluate the very complicated  contour integration at all 
to connect the filling fraction with the unknowns.  

Section \ref{conclusion} is the summary and discussion. 
In appendix  \ref{app}, the details of the proof are found
how the self-consistency of the flow equation 
determines the singular structure of the inner product.

\section{Irregular conformal block} \label{sec:solve}

In this section, we provide explicit results of $Z_N^{(m)}$  for the case $m=1,2,3$.
Similar calculation for  $m=1,2$ can be found in  \cite{NR2012}. 
Explicit expressions of the partition function 
in terms of the parameters of the potential are useful 
to find out the role of parameters in the irregular conformal block. 
The results will be used to check the calculation done 
in the next section using a quite different method.  

\subsection{Loop equation and flow equation}

We briefly review the method of finding the partition function used in \cite{NR2012}.
For simplicity we use the large $N$ limit of the loop equation\cite{BIPZ1983, ACKM1993, A1996}
\begin{equation}
4 W(z)^2 -4V'(z) W(z)=f(z)
\label{loop-eq}
\end{equation}
where $ W(z) =(\hbar b/ 2) 
\left \langle \sum_i  1 /(z - \lambda_i) \right\rangle$ is the resolvent 
and $\vev{\cdots}$ refers to the expectation value with respect to $Z_N^{(m)}$. 
$ f(z) = 2 \hbar b \left\langle 
\sum_i  ( -V'(z)+V '(\lambda_i))/(z - \lambda_i)  \right\rangle $i
is the quantum correction  and is related with the partition function 
\be
f(z)=\sum_{k=0}^{m-1} \frac{v_k(-\hbar^2  \log Z_N^{(m)} )}{z^{2+k}} \,,
~~~~
v_k \equiv   \sum_{s=1}^{m}s c_{s+k}  \frac {\partial }{\partial c_s}  
\ee
where  the potential \eqref{potential-m} is used. 
As the result, we have $m$-coupled differential equations 
in  the parameter space $\{c_\ell\}$
when the loop equation \eqref{loop-eq} is expanded in the 
inverse power of $z$.  
This equation is called the flow equation.
\be
v_k(-\hbar^2  \log Z_N^{(m)})=d_k~~~~{\rm for~} k=1, \cdots, m\,.
\label{flow-eq}
\ee
Here  $d_k$ is the coefficient obtained from the LHS of \eqref{loop-eq}
and is given in terms of the expectation values of the powers of  $\lambda_i$'s. 
Once we find the coefficients  $d_k$  as an  explicit function of $\{c_\ell \}$,
we can find the partition function.  

We will find $d_k$ in an explicit function of $c_\ell$'s 
under the following framework. 
The parameters $c_\ell$ are defined so that the potential  \eqref{potential-m} 
has $m$-distinct saddle points  ($z^{m+1} V'(z)=0$).  
Especially, we assume $c_0 <0$
and  $c_\ell$'s ($\ell \ge 1$ ) are  
alternating in sign in the ascending order in $\ell$ so that $c_1>0$. 
Also we assume a special hierarchical ordering of the parameters 
$ \left|{c_{k+1}}/{c_k}\right| \ll \left|{c_k}/{c_{k-1}}\right|$. 
In terms of parameter $\eta_k = {c_{k+1} c_{k-1}}/{c_k^2}$, the 
hierarchical ordering shows that $|\eta_k| \ll 1$ and 
one may confirm that each saddle point
is proportional to ${c_{k+1}}/{c_k}$ in the leading order of $\eta_{k}$.

In addition, one can demonstrate that 
the hierarchical ordering of the parameters corresponds to the 
special ordering of the position of vertex operators 
of the conformal block. 
To see this let us consider $m=2$ case. 
From the relation \eqref{c_k} one has
$ c_0=\alpha_0 + \alpha_1 + \alpha_2 $, 
$c_1=-\alpha_1 z_1 -\alpha_2 z_2 $ and $c_2= \alpha_0 z_1 z_2 $. 
Its solution has the form $z_a=  (c_1 / \alpha_a)  x_a $ where
$ x_1 =  (1 + \sqrt{1- 4 \eta_1 {\alpha_1 \alpha_2}/(\alpha_0 c_0 )})/2 $ and 
$x_2 =( 1 - \sqrt{1- 4 \eta_1 \alpha_1 \alpha_2 /(\alpha_0 c_0 )})/2 $. 
This shows that if one  assumes  $ (\alpha_1 \alpha_2/\alpha_0)$ is finite and small
at the colliding limit, 
then one can expand $x_a$'s in $\eta_1$ power series assuming $|\eta_1| \ll 1$,
\be
x_1 =1+ \CO( \eta_1) \,,~~~
x_2 =  \eta_1 \frac{\alpha_1 \alpha_2}{\alpha_0 c_0}\left( 1+ \CO(\eta_1)\right) \,.
\ee
It is worth to note that $x_1 = \CO(1)$ and $x_2= \CO( \eta_1)$.
In general, with the scaling of $z_a= c_1 x_a /\alpha_a$, 
one has $x_a$ as $x_a =\CO(\prod_{k=1}^{a-1} \eta_k)$.  
Therefore, the condition $|\eta_k| \ll 1$ is
equivalent  to put the positions of the primary fields with the hierarchy
$|z_m| \ll |z_{m-1}| \ll \cdots \ll |z_2| \ll |z_1| \to 0$.

\subsection{Inner product  $\vev{\Delta|I^{(m)}}$  }

Let us consider the simplest case $m=1$. 
The potential is given as $ V(z) =-c_0  \log(z) +c_1/z$ 
and  has a stable equilibrium point on the positive axis of $z$ 
if $c_0<0$ and $c_1>0$. 
The partition function   $Z_N^{(1)}$  is the function of $c_0$ and $c_1$
and the flow equation\eqref{flow-eq} has the form 
\begin{equation}
c_1 \frac{\partial}{\partial c_1}(-\hbar^2 \log Z_N^{(1)})=d_0 \,.
\label{d1-flow}
\end{equation} 
$d_0$ is simply  obtained from the LHS of \eqref{loop-eq}. 
The result is  $d_0= h_1$ where we use the notation 
$h_1=\hbar b N(\hbar b N+2 c_0)$ for later convenience. 
From this information, one solves the equation \eqref{d1-flow} 
and gets the partition function
$ Z_N^{(1)} (c_0, c_1)  =(c_1) ^{-h_1/\hbar^2} Z (c_0) $ 
where 
$ Z (c_0)$  is independent of $c_1$ but depends on $c_0$. 

The flow equation, however, does not give any information on $c_0$ dependence. 
To see the $c_0$ dependence explicitly, one may rescale the integration variables 
$z =  \xi  c_1  $  in the partition function  $Z_N ^{(1)} (c_0, c_1)  $ 
to put the factorized into the form
$ Z_N ^{(1)} (c_0, c_1)  = (c_1) ^{-h_1/\hbar^2}\, Z(c_0) $ 
where 
\be
Z(c_0)  =  
 \int \prod_i^N d\xi_i \Delta^ {2 \beta} 
e^{ \frac{  \sqrt{\beta}}{g} \sum_i  ( c_0 \log \xi_i - 1/\xi_i )}\,.
\label{Z-c0} 
\ee
If the integration variable is inverted  $\xi_i =  1/x_i$  
this can be put in a more  familiar form 
\be
Z(c_0)  = \int \prod_i^N dx_i~ \Delta^ {2 \beta} 
e^{ \frac{ \sqrt{\beta}}{g} \sum_i  (\alpha_\infty \log x_i - x_i )}\,.
\ee
where $\alpha_\infty = -c_0  - g (2 + 2 \beta(N-1))/\sqrt{\beta}$. 
The integration variable is  $x_i \ge 0$ 
and $\alpha_\infty$ is in the proper range 
so that the partition function is well-defined.
Rescale $x_i $ by $ \alpha_\infty $  and we have  
\be 
Z(c_0)   = (\alpha_\infty)^{h_1/\hbar^2}  \,  \int \prod_i^N dx_i ~
\Delta(x_i) ^ {2 \beta}  e^{ \frac1{g_s} \sum_i 
\Big(    \log(x_i ) -x_i \Big) } 
\label{Z-c0-linear} 
\ee
with $1/ g_s=\alpha_\infty {\sqrt{\beta}  }/ g  $. 
When $\beta=1$, this partition function is the one considered 
by Penner \cite{P1988} to describe the pseudo Euler characteristic.

The partition function $ Z_N ^{(1)} (c_0, c_1)  $ is also obtained from 
the colliding limit of  the 3-point conformal block directly.
The 3-point function is given by
\be
\CF_{3}(\{e^{\alpha_k \phi(z_k)}\}) =|z_0-z_1|^{2 \gamma_3} |z_1-z_\infty|^{2 \gamma_1}
|z_\infty-z_0|^{2 \gamma_2} C (\alpha_0, \alpha_1, \alpha_\infty)
\ee
where $\gamma_1 =\Delta_0-\Delta_1-\Delta_\infty$,
$\gamma_2 =\Delta_1-\Delta_\infty-\Delta_0$,
$\gamma_3 =\Delta_\infty-\Delta_0-\Delta_1$
and $C(\alpha_0, \alpha_1, \alpha_\infty)$ is a $z_k$-independent constant \cite{ZZ,DO}.
Putting $z_0=0$, $z_\infty \to \infty$ and $z_1 =-  c_1/\alpha_0$
and using $\alpha_0+\alpha_1+\alpha_\infty +b N= Q$, one has
\be
(z_1)^{2\alpha_0 \alpha_1} \CF_{3} = 
(c_1) ^{-h_1/\hbar^2} (-\alpha_0) ^{h_1/\hbar^2}  C (\alpha_0, \alpha_1,  \alpha_\infty)
\ee
up to an appropriate normalization.
The power behavior of $c_1$ is the same as  the one obtained from \eqref{d1-flow}
and  $Z(c_0 )$ is identified as
 $Z(c_0 ) =(-\alpha_0) ^{h_1/\hbar^2}   C(\alpha_0, \alpha_1,  \alpha_\infty)$
at the colliding limit $\alpha_0$, $\alpha_1 \to \infty$ 
maintaining $c_0=\alpha_0+\alpha_1$ finite and 
$ \alpha_\infty = Q- bN -c_0$.  
From now on, we do not bother to find the $c_0$ dependence 
which may be  regarded as the normalization of the partition function.  
(Further comment is found in section \ref{conclusion}).

When $m=2$, we have two flow equations.
\begin{gather}
\left(c_1 \frac{\partial}{\partial c_1}+ 2 c_2 \frac{\partial}{\partial c_2}\right) (-\hbar^2 \log Z^{(2)}_N)=d_0
\label{eq:m21} \\
c_2 \frac{\partial}{\partial c_1}(-\hbar^2 \log Z^{(2)}_N) = d_1 \label{eq:m22}
\end{gather}
where $d_0=h_1$ as in the case $m=1$. $d_1=2 \hbar b N c_1+h_2 \langle \sum_{i=1}^{N}\la_i \rangle$ is given in terms of expectation values 
$\langle \lambda_i \rangle$ and  $h_2= 2 \hbar b (\hbar b N+ c_0)$. 
The equation \eqref{eq:m21} forces the partition function of  the form 
\begin{equation}
- \hbar^2 \log Z^{(2)}_N = h_1 \log c_1+H^{(2)}(\eta_1) \label{eq:m23}
\end{equation} 
where $H^{(2)}(\eta_1)$ is the homogeneous solution. 
According to \eqref{eq:m22}, $H^{(2)}(\eta_1)$ obeys 
\be 
2\eta_1^2 \frac{\partial}{\partial \eta_1} H^{(2)}(\eta_1)= h_1 \eta_1- D_1
 \label{eq:m24}
\ee
where $D_1 \equiv { d_1 c_0}/{c_1}$ 
and its parametric dependence on $\eta_1$ is found using the filling fraction $N_k$
\be
\frac{\hbar b N_k}{2}=\oint_{\CA_k} \frac{dz}{2 \pi i} W(z) \,, \quad k=1,2
\label{ff1}
\ee
where $\CA_k$ is the contour loop ($A$-cycle) which includes 
the eigenvalues around the saddle point.
(We freely use the notation $N_k$ for the filling fraction 
instead of the ratio $N_k/N$).
Putting the resolvent $ 2 W(z)= V'(z) + \sqrt{V'(z)^2+f(z)}  $ 
from   \eqref{loop-eq},  
one has parametric relation of the filling fraction 
\be 
\hbar b N_k = \oint_{\mathcal{A}_k} \frac{dz}{2 \pi i}
\frac{\sqrt{ \mathcal{P}_4(z)}}{z^3} 
\ee
where $ 
\mathcal{P}_4(z)=(d_0+c_0^2)z^4+(d_1+2 c_0 c_1)z^3+(c_1^2+2 c_0 c_2)z^2+2c_1 c_2 z+c_2^2$. 
One may assume that $\mathcal{P}_4(z)$ has four real and positive roots 
which need to be justified a posteriori. In this case, there are two branch cuts 
and branch points are the roots of $\CP_4(z)=0$.  
For notational purpose, we denote the saddle point around  $|{c_k}/{c_{k-1}}|$
as  the $k$-th saddle point where the $k$-th cut and the filling fraction $N_k$ 
are associated. 

The filling fraction \eqref{ff1} has a relation $N_1+N_2=N$ 
and therefore, $N_1$ is enough to find $D_1$. 
Integration is given in the elliptic function.
But we will follow a  practical way which works for $m \geq 2$.
Using the integration variable $z= \xi c_1/c_0$, one has 
the rescaled $ P_4(\xi)$ and the first cut is $\CO(\eta_1^0)$.
Therefore it is convenient to put  $P_4(\xi)=\xi^2 \tilde{\CP}_2(\xi)+\CO(\eta_1)$
where $\tilde{\mathcal{P}}_2(\xi)=(d_0+c_0^2)\xi^2+(D_1+2 c_0^2) \xi+c_0^2$.
Therefore, the branch point is given by $\tilde{\CP}_2(\xi)=0$ to the leading order of 
$\eta_1$, 
Expanding in  powers of  $\eta_1$, 
\be
\hbar b N_1 = \oint_{\CA_1} \frac{d \xi}{2 \pi i} \Biggl(\frac{\sqrt{\xi^2 \tilde{\CP}_2(\xi)}}{\xi^3}+\eta_1 \frac{c_0^2 \xi(\xi+1)}{\xi^3 \sqrt{\xi^2 \tilde{\CP}_2(\xi)}}+\mathcal{O}(\eta_1^2)\Biggr) 
\ee
one has the residue integration at $\xi=0$ and $\xi=\infty$. 
\begin{equation}
\hbar b N_1=\hbar b N-\frac{D_1}{2 c_0}+\eta_1 \left( \hbar b N+\frac{(\hbar b N)^2-2D_1}{2 c_0}-\frac{3 D_1^2}{8 c_0^3} \right)+\mathcal{O}(\eta_1^2).
\end{equation}
Finding $D_1$ in small $\eta_1$ expansion, one has 
\begin{equation}
D_1=2 \hbar b N_2 c_0+\eta_1 \left( \hbar^2 b^2(-2 N^2+6 N N_1-3N_1^2)-2 \hbar b c_0(N-2 N_1) \right)+\mathcal{O}(\eta_1^2).
\end{equation} 
Equipped with the explicit $D_1$, we solve the equation (\ref{eq:m24}) 
to find 
\begin{equation}
Z_N^{(2)} = (c_1)^{-h_1/\hbar^2} (\eta_1)^{-\frac{b N_2}{2 } (3 b N_2+4{c_0}/{\hbar})}
e^{-\frac{ b N_2 c_0/\hbar}{ \eta_1} +\mathcal{O}(\eta_1)} \label{eq:m25}
\end{equation}

\begin{figure}
\centering
\includegraphics[width=0.6\textwidth]{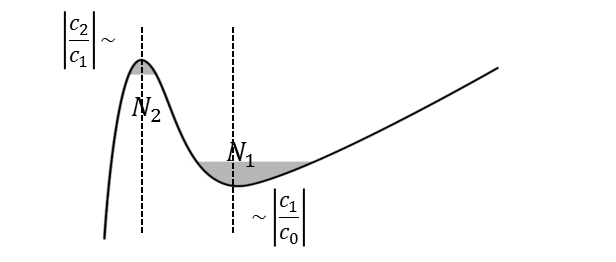}
\caption{Shape of $m=2$ potential}
\label{f:2_2}
\end{figure}

The partition function  $Z_N^{(2)}$ has an interesting feature of the singularity  in $\eta_1$. 
The term with the factor  $N_2^2$ comes from  the Vandermonde determinant. 
On the other hand, the term with factor linear in $N_2$ 
has the essential singularity of  the form $\exp (1/\eta_1) $ 
which is proportional to the energy difference of the two saddle points 
which is the instanton effect  \cite{Kawai2004,MSW2008}.
Suppose we put $N_2=0$ which has no instanton at all. 
Then, the partition function becomes  regular in $\eta_1$, 
$Z_N^{(2)}=(c_1)^{-h_1/\hbar^2} \left(1+\CO(\eta_1)\right)$
and  if one puts $\eta_1 \to 0$, the partition function 
reduces to $Z_N^{(1)}$, the one with $m=1$.
In other words, when  $N_2=0$,  the partition function 
$Z_N^{(2)}$ has the smooth limit  to  $Z_N^{(1)}$
if $ \eta_1 \to 0$ because the singular dependence of $\eta_1$ disappears. 
This limiting procedure is very general
as can be found below  for the case with $m=3$.

When $m=3$, we have three flow equations,
\begin{align}
\biggl(&c_1 \frac{\partial}{\partial c_1}+ 2 c_2 \frac{\partial}{\partial c_2}
+3 c_3 \frac{\partial}{\partial c_3}\biggr) 
(-\hbar^2 \log Z_N^{(3)}) = d_0 \label{eq:m31} \\
\biggl(&c_2 \frac{\partial}{\partial c_1}
+2 c_3 \frac{\partial}{\partial c_2} \biggr)(-\hbar^2 \log Z_N^{(3)}) 
=d_1 \label{eq:m32} \\
&c_3 \frac{\partial}{\partial c_1}(-\hbar^2 \log Z_N^{(3)}) = d_2  \label{eq:m33}
\end{align}
where $d_0 = h_1 $ and $d_1= 2 \hbar b N c_1
+h_2 \langle \sum_{i=1}^{N} \lambda_i \rangle$ 
 as found in \eqref{eq:m21} and  \eqref{eq:m22}.  
$d_2$ is new and 
$ d_2 = 2 \hbar b N c_2+2 \hbar b c_1 \langle \sum_{i=1}^{N}\la_i \rangle
+\hbar^2 b^2 \langle \sum_{i=1}^{N}\la_i \rangle^2
+h_2 \langle \sum_{i=1}^{N}\lambda_i^2 \rangle  $. 

According to \eqref{eq:m31}, one puts the partition function of the form
\begin{equation}
-\hbar^2 \log Z_N^{(3)} = h_1 \log c_1+H^{(3)}(\eta_1,\eta_2)
\end{equation}
where $H^{(3)}(\eta_1, \eta_2)$ is the homogeneous solution 
and is written as the function of 
$\eta_1={c_0 c_2}/{c_1^2}$ and $\eta_2={c_1 c_3}/{c_2^2}$.  
The differential equation of $H^{(3)}(\eta_1, \eta_2)$  is obtained from 
the rest of the equations  (\ref{eq:m32}) and (\ref{eq:m33})
\begin{align}
\frac{\partial H^{(3)}}{\partial \eta_1}&=\frac{D_2-(D_1+4D_2)\eta_2+4d_0 \eta_1 \eta_2^2}{6\eta_1^2 \eta_2^2} 
\label{eq:m36}\\
\frac{\partial H^{(3)}}{\partial \eta_2}&=\frac{D_2-(D_1+D_2)\eta_2+d_0 \eta_1 \eta_2^2}{3\eta_1 \eta_2^3} 
\label{eq:m37}
\end{align}
where $D_1 \equiv d_1  {c_0}/{c_1} $ and $D_2 \equiv  d_2 {c_0}/{c_2} $
are fixed by the filling fraction 
\be
\hbar b N_k = \oint_{\mathcal{A}_k} \frac{dz}{2 \pi i} \sqrt{\frac{\mathcal{P}_6(z)}{z^8}}
\,, \quad k=1,2,3
\ee
where  $N=N_1+N_2+N_3$ and 
$\mathcal{P}_6(z)
=(d_0+c_0^2)z^6+(d_1+2c_0 c_1)z^5+(c_1^2+2c_0 c_2+d_2)z^4 
 +2(c_1 c_2+c_0 c_3) z^3+(c_2^2+2c_1 c_3)z^2+2 c_2 c_3 z+c_3^2 $. 

\begin{figure}
\centering
\includegraphics[width=0.44\textwidth]{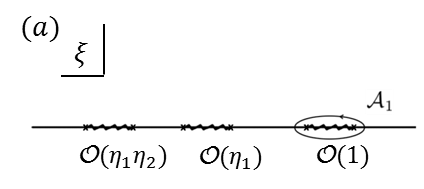}
\quad \qquad
\includegraphics[width=0.44\textwidth]{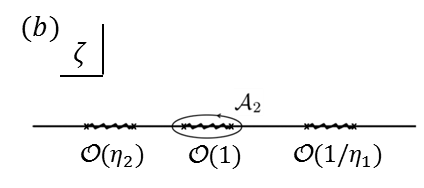}
\caption{Contours for the first(a) and second(b) cut}
\label{f:cut}
\end{figure}

The contour integration $\CA_1$ (figure \ref{f:cut}(a))
is around $z=|{c_1}/{c_0}|$.  
We may rescale $z= \xi {c_1}/{c_0}$ so that
 $({c_0}/{c_1})^6 \CP_6 ( \xi {c_1}/{c_0})
=\xi^4 \tilde{\CP}_2(\xi)+
\eta_1 \xi^3 ((D_2+2 c_0^2) \xi+2 c_0^2 )
+\CO (\eta_1^2 )$  where 
 $ \tilde{\mathcal{P}}_2(\xi)=(d_0+c_0^2)\xi^2+(D_1+2 c_0^2) \xi+c_0^2$. 
Here $\xi$,  $D_1$ and $D_2$ are assumed $\CO(1)$. 
Expanding the filling fraction $N_1$   in powers of   $\eta_1$ and $\eta_2$,
one has 
\be
\hbar b N_1 = \oint_{\mathcal{A}_1} \frac{d \xi}{2 \pi i} \Biggl(\frac{\sqrt{\xi^4 \tilde{\mathcal{P}}_2(\xi)}}{\xi^4}+\eta_1 \frac{(D_2+2c_0^2)\xi+2c_0^2}{2 \xi^3 \sqrt{ \tilde{\mathcal{P}}_2(\xi)}}+\mathcal{O}(\eta_1^2)\Biggr) \,.
\ee
The residues at $\xi=0$ and $\xi=\infty$  give the contour contribution 
\begin{equation}
\hbar b N_1=\hbar b N-\frac{D_1}{2 c_0}+\eta_1 \frac{4c_0^2(h_1-2D_1+D_2)+D_1(2D_2-3D_1)}
{8 c_0^3}+\mathcal{O}(\eta_1^2)\,.
 \label{eq:m34}
\end{equation}

\begin{figure}
\centering
\includegraphics[width=0.5\textwidth]{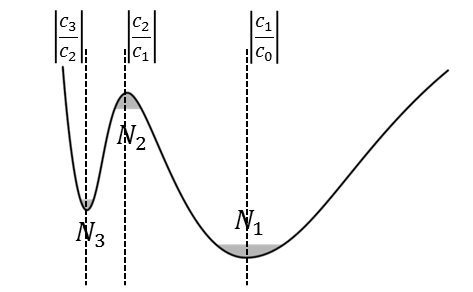}
\caption{Shape of $m=3$ potential}
\label{f:m=3}
\end{figure}

The second cut lies around $z=|c_2/c_1|$.  
We rescale $z=\zeta{c_2}/{c_1} $  for the contour integrations $\CA_2$
(see fig.\ref{f:cut}(b)) and find the polynomial 
$ \eta_1^2 ({c_1}/{c_2})^6  \CP_6( \zeta{c_2}/{c_1})
=c_0^2\zeta^2 (\zeta+1)^2+2 \eta_2 c_0^2 \zeta(\zeta+1)+\eta_2^2 c_0^2 
+\eta_1 \{ \zeta^4 \bigl( (2c_0^2+D_1) \zeta+2c_0^2+D_2 \bigr)+\eta_2 (2 c_0^2 \zeta^3) 
\} +\eta_1^2  \zeta^6 (d_0+c_0^2) $
where  the dominant part is the squared form  $(\zeta^2 +\zeta )^2$.
Therefore, the small $\eta_1$ and $\eta_2$ expansion has no branch cut integral.
After integration, one has
\begin{equation}
\begin{split}
\hbar b N_2&=\frac{1}{2c_0} \left(D_1-D_2+(D_1-2D_2)\eta_2+3(D_1-2D_2)\eta_2^2+\mathcal{O}(\eta_2^3) \right)\\
&+\eta_1 \biggl[ \frac{D_1(3D_1-2D_2)-4 c_0^2(h_1-2D_1+D_2)}{8 c_0^3}\\
&\qquad \qquad \qquad +\frac{3D_2^2-2 c_0^2(h_1-3D_2)}{4 c_0^3}\eta_2^2+\mathcal{O}(\eta_2^3)\biggr]+\mathcal{O}(\eta_1^2).
\end{split} \label{eq:m35}
\end{equation}
$D_1$ and $D_2$ are obtained from equations (\ref{eq:m34}) and (\ref{eq:m35}) 
\begin{align}
D_1&=2 \hbar b c_0 (N-N_1)+ \eta_1 \bigl[A_1+A_2 \eta_2+\mathcal{O}(\eta_2^2) \bigr]
+\mathcal{O}(\eta_1^2,\eta_2^3) \\ 
D_2&= 2 \hbar b c_0 N_3+\eta_2 \left(2 \hbar b c_0(N_2-N_3) \right)+\eta_2^2 \left( 2 \hbar b c_0(N_2-N_3) \right)\\
&+\eta_1 \bigl[A_1\eta_2+B_1 \eta_2^2+\mathcal{O}(\eta_2^3) \bigr]
+\mathcal{O}(\eta_1^3,\eta_2^3)
\end{align}
where
$ A_1= -2 \hbar b c_0 (N_2-N_1) - 2\hbar^2 b^2 N(N_2-N_1)-\hbar^2 b^2 N_1(-2N_2+N_1) $,
$A_2= (-2\hbar b c_0-2\hbar^2 b^2(N-N_1))(N_3-N_2) $, 
and 
$ B_1= 2\hbar b c_0(N_3-N_2)+\hbar^2 b^2(3 N^2+6 N_2^2+10N_1 N_2+3N_1^2-2N(3N_1+5N_2))$. 
Plugging $D_1$ and $D_2$ into a system of equations (\ref{eq:m36}) and (\ref{eq:m37}), we have the partition function
\begin{equation}
\begin{split}
-\hbar^2 \log Z_N^{(3)}&= h_1 \log c_1-\frac{ \hbar b c_0 N_3}{3 \eta_1 \eta_2^2}
+\frac{2 \hbar b c_0 N_3}{\eta_1 \eta_2}+\frac{\hbar b c_0(N_2-N_3)}{\eta_1} \\
&+\frac{\hbar b}{2}(4 c_0 (N-N_1)+\hbar b (3 N_2^2+4N_2 N_3+3N_3^2))\log \eta_1 \\
&+2 \hbar b N_3(\hbar b N_3+c_0) \log \eta_2+\CO(\eta_1,\eta_2)
\end{split}
\ee
or
\be
\begin{split}
Z_N^{(3)}&=( c_1) ^{-h_1/\hbar^2} 
(\eta_1) ^{-\frac{b }2 (N_2(3 bN_2 + 4 c_0/\hbar)
+N_3(3 bN_3 + 4 c_0/\hbar) +4 b N_2 N_3) }
(\eta_2)^{-2bN_3(bN_3 + c_0/\hbar)}   \\
&~~~~ e^{-\frac{b c_0}{\eta_1 \hbar} \left( (N_2-N_3)  
+   \frac{2 N_3}{\eta_2} -\frac{ N_3}{ 3 \eta_2^2}  \right)+\CO(\eta_1,\eta_2)} \,.
\end{split} \label{zn3}
\end{equation} 
Note that the term linear in $N_3$ shows the non-trivial instanton effect
whose exact contribution is not easy to calculate in other ways.
When $N_3 = 0$, the partition function  has the smooth limit  $Z_N^{(m=2)}$ as $\eta_2 \to 0$.

\subsection{Inner product $\langle I^{(n)}| I^{(m)} \rangle$} \label{sec:inim}
\begin{figure}
\centering
\includegraphics[width=0.8\textwidth]{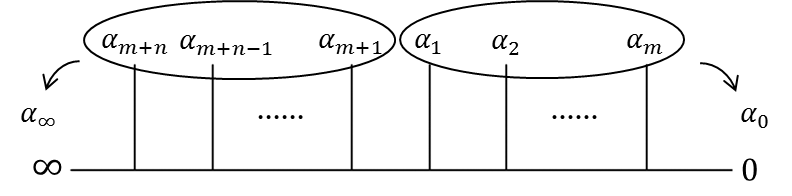}
\caption{Schematic diagram of $\vev{I^{(n)}|I^{(m)}}$}
\label{f:inin}
\end{figure}

Let us now consider  the colliding limit of the conformal block 
as shown in figure \ref{f:inin}.  
In this limit, the partition function  $Z_N^{(n;m)}$ 
has the potential
$V(z)=-c_0 \log z+V_{-}(z)+V_{+}(z)$ 
\be 
V_{-}(z)=\sum_{s=1}^{m} \frac{c_s}{s z^s} \, , \quad V_{+}(z)=-\sum_{t=1}^{n} \frac{c_{-t}z^t}{t}\,.
\label{eq:nm-pot}
\ee
The potential is modified to have a new Liouville momentum  
$c_0=\sum_{s=0}^m \alpha_s$  at zero and  
$c_\infty= \alpha_\infty+ \sum_{t=m+1}^{m+n} \alpha_t$ at infinity
so that  the neutrality condition is $c_0+c_\infty+\hbar b N=\hbar Q$.
The additional positive power term in  $V_{+}(z)$ characterizes 
the irregular singularity at infinity. 
Thus, this matrix model  is identified with the inner product 
$\vev{I^{(n)}|I^{(m)}}$  
between the irregular vector of the  rank $n$ at infinity 
and irregular vector of  the rank $m$ at zero.  

The quantum correction $ f(z)=\sum_{k} {d_k} /{z^{2+k}}$  
has $d_k=0$ when $k \geq m$ or $k \leq -(n+1)$
and $d_{-n}=2\hbar b N c_{-n}$.
The remaining terms are given as the $(m+n-1)$ flow equations 
\begin{align}
&d_k= v_k  (-\hbar^2  \log Z_N^{(n;m)})  \,, ~~~~ 
v_k =   \sum_{s=1}^{m}s c_{s+k}  \frac {\partial }{\partial c_s}~~~
\mathrm{when} ~ 0 \leq k \leq m-1 
\label{inim_dk} \\
&d_{-k}=2 \hbar b N c_{-k}+u_k(-\hbar^2 \log Z_N^{(n;m)}) \,, ~~~
u_k=\sum_{t=1}^n t c_{-t-k} \frac{\partial}{\partial c_{-t}} ~~~
\mathrm{when} ~1 \leq k \leq n-1    \,.
\label{inim_d-k}
\end{align}

If one defines $\varphi(z) \equiv V'(z)^2+f(z)$, one has 
$ \varphi(z)=\sum_{k=-2n}^{2m} {\Lambda_k}/{z^{k+2}}$ 
where $ \Lambda_k=d_k+\sum_{k=\ell+\ell'} c_\ell c_{\ell'}$. 
Regarding $\varphi(z)$ as the expectation value of 
the energy momentum tensor 
$ \varphi(z)={\vev{I^{(n)}|T(z)|I^{(m)}}}/{\vev{I^{(n)}|I^{(m)}}} $, 
one has the eigenstate at origin
\be
L_k |I^{(m)} \rangle= \Lambda_k |I^{(m)} \rangle  \quad
\mathrm{when} ~~ m \leq k \leq 2m
\ee
and another at infinity
\be
\langle I^{(n)} | L_k= \Lambda_k  \langle I^{(n)} |  \quad
\mathrm{when} ~-2 n \leq  k \leq -n \,.
\ee
This definition is consistent with the conjugate of  Virasoro generators
 $L_{-k}=L_k^\dagger$.

We assume $\eta_k \ll1$ for $-(n-1) \leq k \leq m-1$
which ensures that each saddle point is proportional to $|{c_{k+1}}/{c_k}|$.
Suppose we rescale the integral variables $\lambda_i$ as $\la_i {c_0}/ {c_{-1}} $,
the partition function has the form 
\be
Z_N^{(n;m)}= \Big(\frac{c_{-1}} {c_0} \Big) ^{(h_1  /\hbar^2  - bNQ) } 
\int \left[ \prod_{i=1}^{N} d \la_i \right] \Delta(\la_i)^{2 \beta} 
\exp \left(-\frac{\sqrt{\beta}}{g} \sum_i \hat{V}(\la_i) \right)
\ee
where $\hat{V}(z)=-c_0 \log z+V_+ (z {c_0}/{c_{-1}}) 
+V_- ( z {c_0}/{c_{-1}} )$. 
Note that $-b N Q$ power of $c_{-1}$  
is the sub-leading  contribution.
In the large $N$ expansion, we have  
\be
-\hbar^2 \log Z_N^{(n;m)}=-h_1 \log (c_{-1}/c_0)  + H^{(n;m)}( \{ \eta_{k} \})
\label{form}
\ee
where $H^{(n;m)}$ is the  function of $\eta_{k}$'s  because  
the scaled potential has the coefficient 
$c_s  ({c_0}/{c_{-1}} )^{-s}
= c_0 \prod_{k=0}^{s-1} \eta_k^{s-k}$
and $c_{-t} ( {c_0}/{c_{-1}} )^{t}
= c_0 \prod_{j=1}^{t-1} \eta_{-j}^{t-j}$.

The partition function can be evaluated as done in  $\vev{\Delta|I^{(m)}}$. 
To find  $d_k$ as the functions  of $\eta_k$'s
we use the filling fraction relations  ($N=\sum_{k=-(n-1)}^{m} N_k$) 
\begin{equation}
\hbar b N_k = \oint_{\mathcal{A}_k} \frac{dz}{2 \pi i} 
\sqrt{\varphi(z)}
= \oint_{\mathcal{A}_k} \frac{dz}{2 \pi i} \frac{\sqrt{\mathcal{P}_{2(m+n)}(z)}}{z^{m+1}}\,.
\label{eq:ffc}
\end{equation}  
$\CP_{2(m+n)}(z)$ is a polynomial of order $2(m+n)$
and the integrand  has $(m+n)$ cuts. 
Integrating over $\mathcal{A}_k$ is done 
after rescaling  $z$ to $ \xi{c_k}/ {c_{k-1}}$ 
so that the $k$-th cut is $\CO({\{\eta_k^0\}})$ in $\xi$-plane. 
On the other hand, $(k+a)$-th cut with  $a\ge 1$ goes to zero, 
$\CO(\prod_{\ell=0}^{a-1} \eta_{k+\ell})$ and 
$(k-a)$-th cut  to infinity, $\CO(\prod_{\ell=1}^a 1/\eta_{k-\ell})$
as  $\{\eta_j \} \to 0$.

The simplest example is  $n=m=1$. 
It is convenient to put $ Z_N^{(1;1)}$ in the form  \eqref{form}  
whose flow equation is 
\begin{equation}
\eta_0 \frac{\partial}{\partial \eta_0} H^{(1;1)}(\eta_0)=d_0\,. \label{eq:111}
\end{equation}
Filling fraction integral has 
$\mathcal{P}_4(z)
=c_{-1}^2 z^4+2 c_{-1} (\hbar b N+c_0) z^3+(2 c_1 c_{-1}+d_0+c_0^2) z^2
+2 c_1 c_0 z+c_1^2$. 
Rescaling  $z$ to $\xi {c_0}/{c_{-1}} $  we have
$ ({c_{-1}}/{c_0})^2 \CP_4  ( \xi {c_0}/{c_{-1}}) 
=\xi^2 \CP_2(\xi)+\eta_0 \left[ 2c_0^2 \xi(\xi+1) \right]+\eta_0^2 c_0^2$
where 
$\mathcal{P}_2(\xi)=c_0^2 \xi^2+2 c_0(\hbar b N+c_0) z+(d_0+c_0^2)$. 
After this  we have 
$ \hbar b N_0 
= \hbar b N+c_0-\sqrt{d_0+c_0^2}+\mathcal{O}(\eta_0) $
or $ d_0=\hbar b N_1 (\hbar b N_1+2 c_0)+\mathcal{O}(\eta_0)$.
Thus the partition function is given as 
\begin{equation}
-\hbar^2 \log Z_N^{(1;1)} = 
-h_1 \log (c_{-1}/c_0)+\hbar b N_1(\hbar b N_1+2c_0) \log \eta_0+\mathcal{O}(\eta_0).
\end{equation}

When $n=1$ and $m=2$, we have two flow equations 
\begin{align}
\eta_0 \frac{\partial}{\partial \eta_0} H^{(1;2)}(\eta_0, \eta_1) &= d_0
\label{n1m2D0} \\
\left( \eta_0 \eta_1 \frac{\partial}{\partial \eta_0}
- 2 \eta_1^2 \frac{\partial}{\partial \eta_1} \right) H^{(1;2)}(\eta_0, \eta_1) &=D_1
\label{n1m2D1}
\end{align}
where $D_1 \equiv  d_1 {c_0}/{c_1}$. 
Using the polynomial  
$ \mathcal{P}_6(z)=c_{-1}^2z^6+(d_{-1}+2c_{-1}c_0)z^5+(2c_1 c_{-1}+d_0+c_0^2)z^4  +(2c_2 c_{-1}+d_1+2c_1 c_0)z^3+(c_1^2+2c_2 c_0)z^2+2c_1c_2z+c_2^2$,  
%
%
we have 
\begin{align}
  \hbar b N_0 &= \hbar b N+c_0-\sqrt{d_0+c_0^2}+\mathcal{O}(\eta_0 )  \nn\\
\hbar b N_1 &= -\frac{D_1}{2 c_0}-c_0+\sqrt{d_0+c_0^2}
+\eta_1 ({4c_0^2 (d_0-2D_1)-3D_1^2})/({8 c_0^3}) +\mathcal{O}(\eta_1^{\phantom{0}2})
\nn
\end{align}
%
%
%
%
Inverting the relations, we have 
$ d_0=\hbar b (N-N_0) (\hbar b (N-N_0)+2c_0) + \mathcal{O}(\eta_0 ) $
and $ D_1=2 \hbar b c_0 N_2+\eta_1 [\hbar^2 b^2 (N^2-3N_2^2-2N N_0+N_0^2)+2(N_1-N_2)c_0]+\mathcal{O}(\eta_1^{\phantom{0}2}) $. Therefore,   
\begin{align} 
H^{(1;2)}(\eta_0, \eta_1)& =\frac{\hbar b c_0 N_2}{\eta_1}+\hbar b (N_1+N_2) \left( \hbar b (N_1+ N_2)+2c_0 \right) \log \eta_0 \nn\\
& ~~+\frac{\hbar b N_2}{2} \left( 3\hbar b N_2+4 c_0 \right) \log \eta_1
+\CO(\eta_0, \eta_1)
\end{align} 
or 
\be
Z_N^{(1;2)}=\Big({c_{-1}}/{c_0} \Big)^{h_1/\hbar^2}
\Big(\eta_0\Big)^{-b (N_1+N_2) \left( b (N_1+ N_2)+2 \frac{c_0}{\hbar} \right)}
\Big(\eta_1\Big)^{-\frac{b N_2}{2} \left( 3 b N_2+4 \frac{c_0}\hbar \right) }
e^{-\frac{b N_2 c_0/\hbar}{\eta_1}+\CO(\eta_0, \eta_1)} \,.
\label{I1I2part}
\ee 

As $m$ and $n$ increase, the flow equations becomes more  complicated. 
We give the explicit result for $n=m=2$ for later comparison.
We have three flow equations 
\begin{align}
\frac{\partial H^{(2;2)} }{\partial \eta_0}&= \frac{d_0}{\eta_0}\\
\frac{\partial H^{(2;2)} }{\partial \eta_1}&=\frac{d_0 \eta_1 -D_1}{2\eta_1^2} \\
\frac{\partial H^{(2;2)} }{\partial \eta_{-1}}&=\frac{d_0 \eta_{-1}-D_{-1}-h_1 \eta_{-1}+2 \hbar b N c_0}{2 \eta_{-1}^{\phantom{-}2}}.
\label{eq:221}
\end{align}
where $D_{-1} \equiv d_{-1} {c_0}/{c_{-1}} $. 
Using the polynomial 
\begin{align}
\begin{split}
\mathcal{P}_8(z)&=c_{-2}^2z^8+2 c_{-1} c_{-2} z^7+(c_{-1}^{\phantom{1}2}+2c_{-2} 
(\hbar b N+c_0)) z^6+(2c_1 c_{-2}+d_{-1}+2 c_{-1} c_0)z^5 \\
&+(2c_1 c_{-1}+2c_2 c_{-2}+d_0+c_0^2) z^4+(2 c_2 c_{-1}+d_1+2 c_1 c_0) z^3 \\
&+(c_1^2+2c_2 c_0) z^2+2 c_1 c_2 z+c_2^2  \nn
\end{split}
\end{align}
the filling fractions are evaluated 
\begin{align}
\hbar b N_0 & =\frac{D_{-1}}{2 c_0}+c_0-\sqrt{d_0+c_0^2} \nn\\
&~~
+\frac{-4 c_0^2(d_0-2 D_{-1})+3D_{-1}^{\phantom{1}2}-4 \hbar b N D_{-1} c_0-8 \hbar b N c_0^3}{8 c_0^3} \eta_{-1}
+\CO(\eta_0, \eta_{-1}^2)  \,, \\
\hbar b N_1
&=-\frac{D_1}{2 c_0}-c_0+\sqrt{d_0+c_0^2}+\frac{4 c_0^2(d_0-2D_1)-3D_1^2}{8 c_0^3} \eta_1
+\CO(\eta_0, \eta_1^2)\\
\hbar b N_2
& =\frac{D_1}{2 c_0}+\frac{-4 c_0^2 (d_0-2D_1)+3D_1^2}{8 c_0^3}+\CO(\eta_0, \eta_1^2) \,.
\end{align}
Inverting the result, one has 
$d_0=A_1$ 
$ D_1=B_1+B_2 \eta_1 $ and $ D_{-1}=C_1+C_2 \eta_{-1}$ 
where
\begin{align}
A_1=&\hbar b (N_1+N_2)(\hbar b(N_1+N_2)+2 c_0), ~~ 
B_1=2 \hbar b c_0 N_2, ~~
C_1=2 \hbar b c_0 (N_0+N_1+N_2) \nn\\
B_2=&\hbar b \left[ \hbar b(N_1^2+2 N_1 N_2-2N_2^2)+2(N_1-N_2)c_0 \right] \nn\\
C_2=&\hbar b \bigl[ 2\hbar b N(N-N_{-1})-3 \hbar b (N-N_{-1})^2 \nn\\
&\qquad+(N_1+N_2)(\hbar b (N_1+N_2)-2c_0)-4 c_0(N-N_{-1})+2 N c_0 \bigr]  \nn
\end{align}
and one has the partition function 
\be
\begin{split}
H^{(2;2)}(\eta_{-1},&\eta_0,\eta_1)=-\frac{\hbar b c_0 N_{-1}}{\eta_{-1}}
+\frac{\hbar b c_0 N_2}{\eta_1}
+\hbar b (N_1+N_2)(\hbar b (N_1+N_2)+2c_0) \log \eta_0 \\
&+\frac{\hbar b N_2(3 \hbar b N_2+4c_0)}{2} \log \eta_1
-\frac{\hbar b N_{-1}(-3\hbar b N_{-1}+4(c_0+\hbar b N))}{2} \log \eta_{-1} \\
&+\CO(\eta_{-1}, \eta_0, \eta_1) 
\end{split}
\ee
or 
\be
\begin{split}
Z_N^{(2;2)}=\Big({c_{-1}}/{c_0} \Big)^{h_1/\hbar^2}& 
\Big(\eta_0\Big)^{-b (N_1+N_2)( b (N_1+N_2)+2 {c_0}/\hbar)}
\Big(\eta_{-1}\Big)^{\frac{ b N_{-1}}2 (-3 b N_{-1}+4({c_0}/\hbar+ b N))} 
\\ &~~~~~~~~~~
\Big(\eta_1\Big)^{-\frac{ b N_2}2(3  b N_2+4{c_0}/\hbar)}
e^{-\frac{ b N_2 c_0 /\hbar}{\eta_1}+\frac{b  N_{-1} c_0/\hbar}{\eta_{-1}}
+\CO(\eta_{-1}, \eta_0, \eta_1)} \,.
\end{split}\label{I2I2part}
\ee


\section{Hierarchical structure of the partition function} \label{sec:hier}
It should be noted that given $N_2$,  the inner products  $\vev{I^{(1)}|I^{(2)}}$ in  \eqref{I1I2part}, 
$\vev{I^{(2)}|I^{(2)}}$ in  \eqref{I2I2part}  and  $\vev{\Delta|I^{(2)}}$
in  (\ref{eq:m25}) have the same  $\eta_1$ dependence. 
Likewise given $N_1$ and $N_2$, 
one can see the same $\eta_0$ dependence in $\vev{I^{(2)}|I^{(2)}}$ and $\vev{I^{(1)}|I^{(2)}}$.
From this observation,  one may wonder  if how much singular structure 
in $\vev{I^{(n)}|I^{(m)}}$ is shared with others. 
We will investigate the possibility of this singular structure in detail 
and present a way  to find the partition function 
without using the filling fraction contour integration.

\subsection{Singular structure in  $\langle I^{(n)} | I^{(m)} \rangle$ and $\langle \Delta | I^{(m)} \rangle$}
Let us investigate the small $\eta_0$-behavior
in the inner product $\vev{I^{(n)}|I^{(m)}}$.
First we rearrange the integration variables in two groups. 
One group is the integration variables which lies around the  $k$-th saddle point
with $k \geq 1$ whose integration variables are denoted  as $\lambda_i^{L}>0$.
The  number of eigenvalues is  $N_L$.
The other group is denoted as $\lambda_i^{R}>0$
which  lives around the $k$-th saddle point with $k \leq 0$.
The number of  $\lambda_i^{R}$ is $N_R$. 
The partition function $Z_N^{(n;m)}$ is rewritten in terms of  two regrouped variables 
\begin{align}
Z_N^{(n;m)} 
&=\int \left[ \prod_{i=1}^{N_L} 
d \lambda_i^{L}  \right] 
\left[ \prod_{j=1}^{N_R} d \lambda_j^{R} \right] 
\Delta  (\lambda^{L})^{2 \beta} \,
\Delta  (\lambda^{R}  )^{2 \beta} 
\prod_{i=1}^{N_L}  \prod_{j=1}^{N_R}
( \lambda_j^{R} -\lambda_i^{L})^{2 \beta} 
 \nn\\
&~~~~~~~~~\times 
 \exp  \left [  - \frac{\sqrt{\beta}} {g} 
\Big( \sum_{i=1}^{N_L} V( \lambda_i^L) + \sum_{j=1}^{N_R} V( \lambda_j^R)
 \Big) \right]
\label{splitpartition} 
\end{align} 
whose potential is given  in \eqref{eq:nm-pot}.
If we use  $\eta_0 \ll 1$,  
we may put $Z_N^{(n;m)}$ into the factorized  form 
$Z_L Z_R \left(1+\CO(\eta_0) \right)$.
The reason is as follows.  
The integration variable $\la_i$ around the $k$-th saddle 
point has the scaling as ${c_{k}}/{c_{k-1}}$.  
Suppose we rescale the  integration variables 
$\la^L_i$ using  the largest scale $ {c_1}/{c_0}$ 
and  $\la^R_i$ using the smallest scale $ {c_0}/{c_{-1}}$.
Then   the ratio $  \la^L/ \la^R $ will be the order of 
$\eta_0 =(c_1 c_{-1})/c_0^2 \ll 1 $. 
Therefore, the dominant  contribution  of the determinant  is 
\be
\prod_{i=1}^{N_L} \prod_{j=1}^{N_R} \left( \lambda_j^{R}-\lambda_i^{L } \right)^{2 \beta} 
=\prod_{j=1}^{N_R} (\lambda_j^R)^{2 \beta N_L}
\Big( 1+\CO (\eta_0)  \Big). 
\label{Vander}
\ee

In addition,  with the rescaling of  $\la^L=\frac{c_1}{c_0} \xi^L$ 
the potential $ V_{+} ( \la^L)$ is put into the form 
$ V_{+} ( \xi^L\, {c_1}/{c_0} )=-\sum_{t=1}^{n}(\eta_0  \,\xi^L )^t
( \prod_{j=1}^{t-1} \eta_{-j}^{t-j} )/t  $
which vanishes as $\eta_0 \to 0$
and is neglected.  
On the other hand,
$ V_{-} (\xi^L{c_1}/{c_0} )
=\sum_{s=1}^{m}  (\prod_{a=0}^{s-1} \eta_{s-a}^{a})/ (s \bigl(\xi^L\bigr)^s) $
is finite as $\eta_0 \to 0$. 
As the result,  the partition function of $\la^L$ has the form 
\be
Z_L=  \int \biggl[ \prod_{i=1}^{N_L} d \lambda_i^{L} \biggr]
 \Delta \bigl(\lambda^{L} \bigr)^{2 \beta} 
\textrm{exp} \biggl[ -\frac{\sqrt{\beta}}{g} 
 \sum_{i=1}^{N_L} V_L ( \lambda_i^L) \biggr] 
\label{lambdaL}
\ee
where $V_L(z)=-c_0 \log z+V_{-}(z)$. Therefore, $Z_L$ is identified as 
the inner product $\langle \Delta(c_\infty+\hbar b N_R) | I^{(m)} \rangle$
where the Liouville  momentum at infinity is $c_\infty+\hbar b N_R$.

Likewise, $\la^R$ contribution is given by
\be 
Z_R=\int \biggl[ \prod_{j=1}^{N_R} d \lambda_j^{R} \biggr]
\Delta \bigl(\lambda^{R} \bigr)^{2 \beta}
\textrm{exp} \biggl[ -\frac{\sqrt{\beta}}{g}\sum_{j=1}^{N_R} V_R(\la_j^R) \biggr] 
\ee
where $V_R(z)=-(c_0+\hbar b N_L) \log z+V_{+}(z)$. 
Here, the potential $V_R (z)$ is not  $- c_0\log z+ V_+(z)$  
but an extra term $-\hbar b N_L \log z$ is added
due to the extra contribution of the Vandermode determinant
\eqref{Vander}. 
It is a simple exercise to show that $V_{-}(\xi^R {c_0}/{c_{-1}}) \to 0 $ 
as $\eta_0 \to0$. 
After this consideration, one notices that 
$Z_R$ is identified with  $\langle I^{(n)} | \Delta (c_0+\hbar b N_L) \rangle$. 

Finally, the sub-dominant part of the determinant in \eqref{Vander} 
has the leading term 
\be
-2\beta Z_{L}Z_{R} \Biggl[ \sum_{i=1}^{N_L} \sum_{j=1}^{N_R}
\left \langle \la_i^L \right \rangle_L  \left\langle\frac{1}{\la_j^R} \right\rangle_R
\Biggr]
\ee
where $\vev{\cdots}_L$ and $\vev{\cdots}_R$ refer to the expectation value
\be
\vev{\CO}_A= \frac{1}{Z_A} \int \biggl[ \prod_{i=1}^{N_A} d \lambda_i^{A} \biggr]
 \Delta \bigl(\lambda^{A} \bigr)^{2 \beta} \CO \, \,
e^{-\frac{\sqrt{\beta}}{g}  \sum_{i=1}^{N_A} V_A( \lambda_i^A)} \,, \qquad A=L, R \,.
\ee
Since the expectation value $\left \langle \la_i^L \right \rangle_L$ is $\CO(c_1/c_0)$
and $\left\langle 1/\la_j^R \right\rangle_R \sim \CO(c_{-1}/c_0)$,
its product is the order of $\eta_0$ and vanishes as $\eta_0 \to 0$.
The next leading contributions should vanish as the high power of $\eta_0$'s.

\begin{figure}
\centering
\includegraphics[width=1\textwidth]{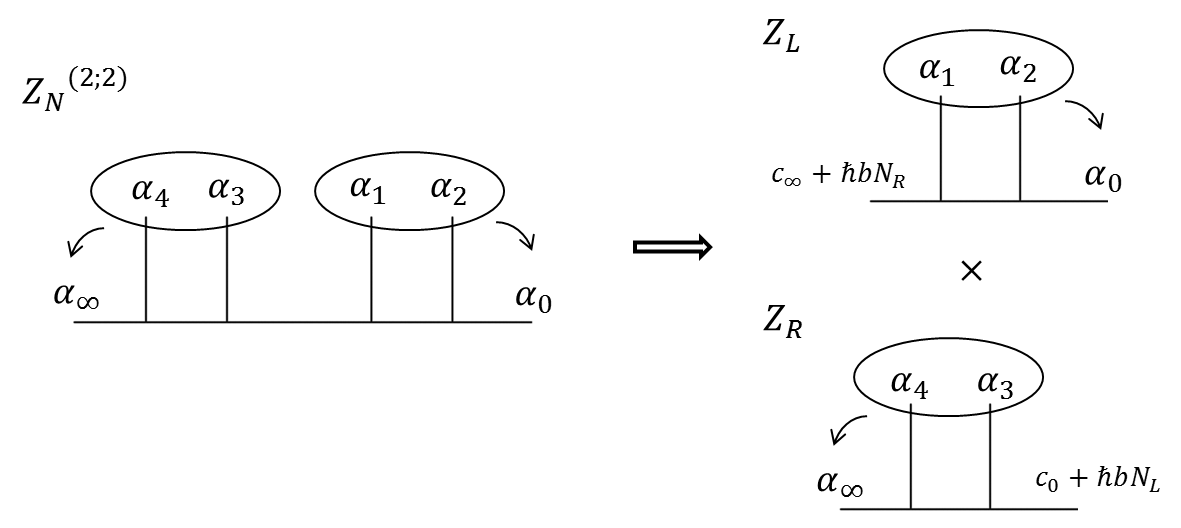}
\caption{The case $m=n=2$. To the zeroth order of $\eta_0$, the partition function is factorized.}
\end{figure}

Thus, one can conclude that 
$Z_N^{(n;m)}=Z_{L} Z_{R} \left(1+\CO(\eta_0) \right)$ and
\be
\vev{I^{(n)}|I^{(m)}} = \vev{I^{(n)} | \Delta (c_0+\hbar b N_L)}
\vev{\Delta (c_\infty+\hbar b N_R)|I^{(m)}}\left(1+\CO(\eta_0) \right) \,.
\label{irr-factor}
\ee
This conclusion is checked by the explicit result given in the previous section.
Since $\langle \Delta | I^{(2)} \rangle =\langle I^{(2)}| \Delta \rangle $ in  (\ref{eq:m25}),
we have 
the inner product $Z_{L}= \vev{\Delta (c_\infty+\hbar b N_R)|I^{(m)}}$ 
\begin{equation}\label{eq:nl} 
-\hbar^2 \log Z_{L} =\hbar b N_L(\hbar b N_L+2c_0) \log c_1+\frac{\hbar b N_2 c_0}{\eta_1}
+\frac{\hbar b N_2}{2}(3 \hbar b N_2+4c_0) \log \eta_1+\CO(\eta_1)
\end{equation}
where $N_L=N_1+N_2$ and $N_R=N_0+N_{-1}$.
And the inner product  $Z_R= \langle I^{(2)} | \Delta (c_0+\hbar b N_L) \rangle$ 
is given as 
\begin{align}
-\hbar^2 \log Z_{R}=&-\hbar b N_R (\hbar b N_R+2 (c_0+\hbar b N_L)) \log c_{-1}
-\frac{\hbar b N_{-1} c_0}{\eta_{-1}} \\
&-\frac{\hbar b N_{-1}}{2}(-3\hbar b N_{-1}+4 (c_0+\hbar b N)) \log \eta_{-1}
+\CO(\eta_{-1})
\end{align}
where the neutrality condition $c_\infty=-c_0-\hbar b N$
is used.  Therefore, the product of $Z_L$ and $Z_R$  is given as  
\begin{equation}
\begin{split}
-\hbar^2 \log (Z_{L} Z_{R})&=-\hbar b N(\hbar b N+2c_0) \log c_{-1}
-\frac{\hbar b c_0 N_{-1}}{\eta_{-1}}
+\frac{\hbar b c_0 N_2}{\eta_1}\\
&+\hbar b N_L(\hbar b N_L+2c_0) \log \eta_0
+\frac{\hbar b N_2(3 \hbar b N_2+4c_0)}{2} \log \eta_1 \\
&-\frac{\hbar b N_{-1}(-3\hbar b N_{-1}+4(c_0+\hbar b N))}{2} \log \eta_{-1}
+\CO(\eta_{-1}, \eta_0, \eta_1)
\end{split}
\end{equation}
where $N=N_L+N_R$.   
This result $Z_{L} Z_{R}$ is in perfect agreement with the inner product 
$\langle I^{(2)} | I^{(2)} \rangle$ shown  in  \eqref{I2I2part}.

\subsection{Hierarchical relation between $\langle \Delta | I^{(m)} \rangle$ and $\langle \Delta | I^{(m-1)} \rangle$}

The singular contribution of $\vev{I^{(n)}|I^{(m)}}$ at  small $\eta_0$ limit  
is given as the product of  $\vev{I^{(n)}|\Delta}$ and $\vev{\Delta|I^{(m)}}$. 
What will happen to the other parameters? 

Let us consider $\vev{\Delta|I^{(m)}}$.  
The  potential has $m$ saddle points, 
$V(z)=-c_0 \log z+V_{-} (z)$ in \eqref{eq:nm-pot}.
Let us concentrate on the integration variable $\la_i^{(m)}$ 
around the $m$-th saddle point
whose  index $i$ running from 1 to $N_m$. 
The other variables  $\la_J$  
has the index  $J$ running from 1 to $\bar{N}_m$ 
so that  $\bar{N}_m + N_m =N $.  

The $m$-th saddle point is much smaller than other saddle points 
 $\la_i^{(m)} \ll  \la_J $ and  therefore,  $\eta_{m-1} \ll 1$. 
The determinant part $\prod_{J=1}^{\bar{N}_m} \prod_{i=1}^{N_m} 
\bigl( \lambda_J-\lambda_i^{(m)} \bigr)^{2 \beta}$
has the dominant contribution 
$\prod_{J=1}^{\bar{N}_m} (\lambda_J)^{2 \beta N_m}$. 
The  partition function  with $\la_J$'s only is $Z^{(m-1)}$:
\be 
Z^{(m-1)} \equiv \int \biggl[ \prod_{J} d \lambda_J \biggr]
\Delta \bigl(\lambda_J \bigr)^{2 \beta}
\textrm{exp} \biggl[ -\frac{\sqrt{\beta}}{g}\sum_{J} V_{m-1} ( \lambda_J) \biggr] \,.
\label{lambdar} 
\ee
The effective potential $ V_{m-1}(z)  $ contains 
$-\hbar b N_m  \log z$  from the determinant part. 
In addition, as  $\eta_{m-1} \to 0$,  
the original term $c_m/z^m$ in the  potential $V(z)$  
drops out.   This is  easily seen  if one scales 
$\la_J$ by  ${c_k}/{c_{k-1}} $  for any  $k \leq m-1$. 
While  $\xi_J  = \CO(\eta_{m-1}^{~0})$, the term $c_m/z^m$ 
is $ \CO(\eta_{m-1})$.  Other terms are finite.  
Therefore we have the effective potential 
\be
V_{m-1} (z)=-(c_0+\hbar b N_m) \log z +\sum_{s=1}^{m-1} \frac{c_s}{s z^s} \,.
\ee  
The partition function $Z^{(m-1)}$ is the inner product $\vev{\Delta|I^{(m-1)} (c_0+\hbar b N_m)}$
and the irregular vector $|I^{(m-1)} (c_0+\hbar b N_m) \rangle$ 
has the rank $m-1$ with the Liouville momentum $\alpha =c_0+\hbar b N_m$. 

The partition function written in terms of $\la_i^{(m)}$ has the form 
\be
T^{(m)} \equiv 
\int \biggl[ \prod_i d \lambda_i^{(m)} \biggr]
 \Delta \bigl(\lambda^{(m)} \bigr)^{2 \beta} 
\textrm{exp} \biggl[ -\frac{\sqrt{\beta}}{g}  \sum_i V ( \lambda_i^{(m)}) \biggr] \,.
\label{lambdal}
\ee
Here, the original potential $V(z)$ is used 
and becomes  infinite as $\eta_{m-1} \to 0$. 
This shows that $T^{(m)}$ contains the singular contribution.

It is noted that  the sub-leading contribution 
in the determinant vanishes  as $\eta_{m-1} \to 0$.  
To check this, let us consider the  $\CO (\la_i^{(m)}/\la_J) $ 
contribution 
\be
Z^{(m-1)} T^{(m)} \Biggl[ \sum_{i=1}^{N_m} \sum_{J=1}^{\bar{N}_m}
\left \langle \la_i^{(m)} \right \rangle_{T^{(m)}} \left\langle\frac{1}{\la_J} \right\rangle_{Z^{(m-1)}} \Biggr]\,.
\ee 
The expectation value $\langle \la_i^{(m)}  \rangle_{T^{(m)}}$ 
is $\CO(c_{m-2}/c_{m-1})$ and 
$\left\langle 1/\la_J \right\rangle_{Z^{(m-1)}}$ 
is $\CO(c_{m}/c_{m-1})$. 
Therefore, its product is the order of $\eta_{m-1}$. 
The higher order contribution  is given as  the higher power of $\eta_{m-1}$'s.

\begin{figure}
\centering
\includegraphics[width=1\textwidth]{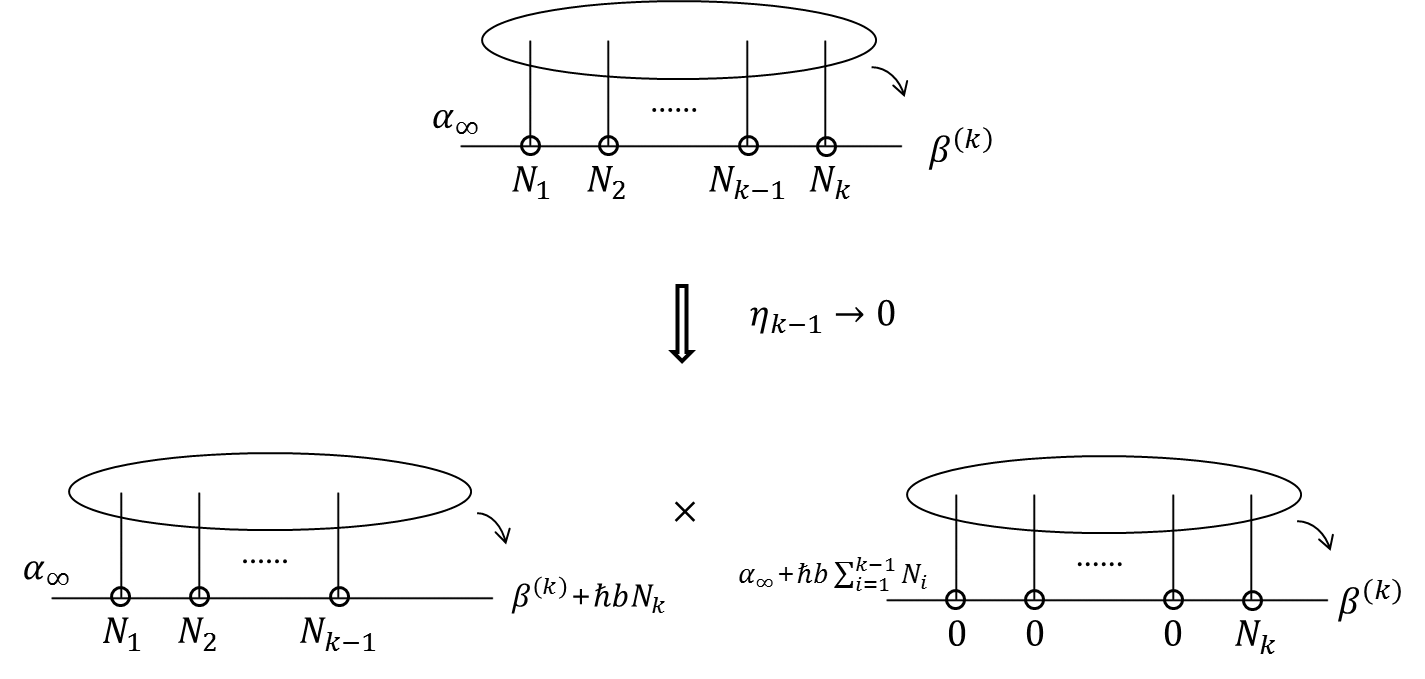}
\caption{Hierarchical structure of $\vev{\Delta|I^{(k)} (\beta^{(k)})}$ as $\eta_{k-1} \to 0$.}
\end{figure}

Considering all the contributions, one concludes that  
$\vev{\Delta|I^{(m)}}=Z^{(m-1)} T^{(m)} \times \left(1+ \CO(\eta_{m-1})  \right)$
which shows that 
\be
\vev{\Delta|I^{(m)}(c_0)}=T^{(m)}(c_0 ; N_m) 
 \vev{\Delta|I^{(m-1)}(c_0+\hbar b N_m)} (1+\CO(\eta_{m-1}))\,.
\label{success}
\ee
In the  small $\eta_{m-1}$ limit,  
the irregular vector of rank $m$ 
is reduced to the one of the  rank $m-1$ 
with the momentum shift $c_0 \to c_0+\hbar b N_m$
and its non-trivial ratio $T^{(m)}(c_0 ; N_m)$ 
contains all the singular contribution. 
When $m=1$, one has  the singular structure of 
$\vev{\Delta|I^{(1)}(c_0)} $ and 
$T^{(1)}(c_0 ; N_1)$ is the same.

One may apply this result \eqref{success} successively
to get the singular part of $\vev{ \Delta| I^{(m)}(c_0)}$  
when the set of parameters $\{ \eta_1, \cdots, \eta_{m-1} \}$ are small.
As the result we have  the following hierarchical structure of the singularity. 
\be
\vev{\Delta|I^{(m)}(c_0)}_S 
=\prod_{k=1}^{m} T^{(k)} ( \beta^{(k)} ; N_k  ) 
 \label{part_hier} 
\ee
where the subscript $S$ stands for the singular part only
neglecting regular contributions.
Here $T^{(k)} (\beta^{(k)} ; N_k )$ is defined in \eqref{lambdal} 
which has filling fraction $N_k$ and Liouiville momentum 
$\beta^{(k)}$ satisfies  the relation 
 $\beta^{(k-1)}=\beta^{(k)}+\hbar b N_k$ and $\beta^{(m)} = c_0$.  

The explicit expression of  $T^{(k)} (\beta; N_k)$ 
can be found from the expressions  in section \ref{sec:solve}.
For example, using the inner product $\vev{\Delta|I^{(2)} (c_0)}$ 
given  in \eqref{eq:m25}, one has $T^{(2)}(c_0;N_2)$ 
by putting $N_1 \to 0$ 
\be
T^{(2)}(c_0;N_2)=(c_1)^{-b N_2 (b N_2+2\frac{c_0}{\hbar})}
 (\eta_1)^{-\frac{b N_2}{2} ( 3 b N_2+4\frac{c_0}{\hbar})}
e^{-\frac{ b N_2 c_0/\hbar}{ \eta_1}+\CO(\eta_1)} \,. \label{sec3t2}
\ee
When $m=3$,  $\vev{\Delta|I^{(3)} (c_0)}$ in \eqref{zn3}  reduces to  
$T^{(3)}(c_0;N_3)$ if $N_1=N_2 \to 0$ 
\be
\begin{split}
T^{(3)}(c_0;N_3)=&(c_1)^{-b N_3 (b N_3+2\frac{c_0}{\hbar})}
 (\eta_1)^{-\frac{b N_3}{2} ( 3 b N_3+4\frac{c_0}{\hbar})} \\
&~~~~~~~
(\eta_2)^{-2 b N_3 ( b N_3+\frac{c_0}{\hbar})}
e^{-\frac{b c_0}{\eta_1 \hbar} \left( -N_3
+   \frac{2 N_3}{\eta_2} -\frac{ N_3}{ 3 \eta_2^2}  \right)+\CO(\eta_1, \eta_2)} \,. \label{sec3t3}
\end{split}
\ee

We note that the singular structure in  \eqref{part_hier} 
is consistent with the ansatz proposed  in \cite{GT2012}. 
When $k=2$, the ansatz is given as
\be
|I^{(2)} (\beta^{(2)}) \rangle
=c_2^{\nu_2} c_1^{\nu_1} e^{(\beta^{(2)}-\beta^{(1)})\frac{c_1^2}{c_2}}
\sum_{j=0}^\infty c_2^j ~ | I_{2j}^{(1)} (\beta^{(1)}) \rangle
\ee
where the vector $| I_{2j}^{(1)} (\beta^{(1)}) \rangle$
is so-called generalized descendants of the rank $1$ irregular vector 
and is the linear combination of vectors obtained by
acting Virasoro generators or $c_1$-derivatives on
$| I_0^{(1)} (\beta^{(1)}) \rangle \equiv | I^{(1)} (\beta^{(1)}) \rangle$.
From the Ward identities the factors were determined as
$\nu_1=2(\beta^{(2)}-\beta^{(1)})(Q-\beta^{(1)})$,
$\nu_2=(\beta^{(1)}-\beta^{(2)})(\frac3 2 Q-\frac 3 2 \beta^{(1)}-\frac 1 2 \beta^{(2)})$.
The prefactor in front of $| I_0^{(1)} (\beta^{(1)}) \rangle$ 
is  $T^{(2)}(\beta^{(2)};N_2)$. ($Q=0$ if the large $N$ limit is taken.) 

\subsection{Flow equations for $T^{(k)} (\beta^{(k)};N_k)$}

It is shown that the singular structure of the inner product between 
the irregular vectors are encoded 
in the partition function $T^{(k)}$ as  in \eqref{part_hier}.
Therefore, if we find $T^{(k)}$, then all the singular structures 
in the parameter space are known.  
In this subsection, we are presenting  a set of differential equations 
for  $T^{(k)}$ so that one can find the partition function 
directly using the flow equations.  
 
We start with the partition function 
\be 
T^{(k)} (\beta^{(k)} ; N_k) =
\int \biggl[ \prod_{i=1}^{N_k} d \la_i \biggr] \Delta (\la)^{2 \beta} 
\textrm{exp} \biggl[ -\frac{\sqrt{\beta}}{g}  \sum_{i=1}^{N_k} 
{V} (\la_i) \biggr] 
\label{Tk}
\ee
where we use the potential 
$ V (z) = -\beta^{(k)} \log z + \sum_{t=1}^k   c_t/(t z^t)$. 
If one scale $\la_i^{(k)}$ by $ {c_k}/{c_{k-1}} $, 
one has 
$ T^{(k)}(\beta^{(k)} ; N_k)=
({c_k}/{c_{k-1}} )^{N_k-b^2 N_k(N_k-1)-2 \beta^{(k)} {b}N_k/{\hbar} } 
~\tilde{T}^{(k)}$ 
where $\tilde{T}^{(k)}$  is given in terms of the rescaled 
integration variables $\xi_i$. Its potential is given as 
\be
\tilde V (\xi ) = \frac{\beta^{(k)}}{ \kappa_0 } 
 \left( -\kappa_0   \log \xi + \sum_{t=1}^k  \frac{ \kappa_t}{t \xi^t} \right) \,.
\ee 
The parameters $\{ \eta_1, \cdots, \eta_{k-1} \}$ are replaced with 
$ \kappa_t \equiv \prod_{\ell=1}^{k-1-t} \eta_{t+\ell}^{~\ell}  $  
for  $(0 \leq t \leq k-2)$ and $\kappa_{k-1}=\kappa_k \equiv 1$. 
The overall parameter $\kappa_0$ is specially treated 
and is equivalently called  $\tau_k$. 
These new parameters have a definite  ordering: 
$\tau_k \ll \kappa_1 \ll \cdots \ll \kappa_{k-2} \ll 1$.
In this rescaling, $ \xi_i $ around the $k$-th saddle point 
and has $\vev{\xi_i^s }=(-1)^s +\CO(\{ \kappa_\ell \})$. 

Defining $\tilde{H}^{(k)}( \{\kappa_\ell\}) \equiv -\hbar^2 \log \tilde{T}^{(k)}$, 
one  has  $(k-1)$-set of differential equations  
\begin{align}
&\frac{\partial \tilde{H}^{(k)}}{\partial \tau_k}
=\frac{2\hbar b \beta^{(k)}}{\tau_k^2} \sum_{t=1}^{k} 
\frac{\kappa_t}t 
\left\langle \sum_{i=1}^{N_k}  \frac 1 { \xi_i^{~t}} \right\rangle
\\ 
&\frac{\partial \tilde{H}^{(k)}}{\partial \kappa_t}
=-\frac{2 \hbar b \beta^{(k)}}{t \tau_k} 
\left\langle \sum_{i=1}^{N_k} \frac{1}{\xi_i^{~t}} \right\rangle 
\,, ~~(1 \leq t \leq k-2)
\end{align} 
Since $\vev{1/\xi_i^t} =  \CO(\{\tau_k ^0 \})$, we may put 
the flow equation in power series of $\tau_k$ as 
\be
\frac{\partial \tilde{H}^{(k)}}{\partial \tau_k}=
-\frac{H^{(k)}_{-1}}{ \tau_k^{2}}+\frac{H_0^{(k)}} {\tau_k}+
\sum_{n \ge 1} n H^{(k)}_n \tau_k^{n-1}
\,, ~~~~~
\frac{\partial \tilde{H}^{(k)}}{\partial \kappa_t}=
\sum_{n \ge -1} G^{(k)}_{t, n} \tau_k^n 
\label{Hkflow}
\ee
where $H_n^{(k)}$ and $G_{t, n}^{(k)}$  are  independent of $\tau_k$ 
and  regular in   $\{\kappa_1, \cdots, \kappa_{k-2} \}$. 
Therefore,  if one integrates \eqref{Hkflow} over  $\tau_k$
one ends up with the form 
\be
\tilde{H}^{(k)}(\{\kappa_\ell\}) =
\frac{H^{(k)}_{-1}}{\tau_k}+H_0^{(k)} \log \tau_k+
\sum_{n \ge 1} H^{(k)}_n \tau_k^n\,.
\label{formofH}
\ee
This shows that  the singular contribution to $T^{(k)}$ as $\eta_\ell \to 0$
is due to the terms  $H_{-1}^{(k)}$, $H_{0}^{(k)}$. 

\subsection{Evaluation of $T^{(k)}$ from the flow equation}

The direct calculation of the partition function
uses the loop equation
and requires the complicated integration
to find the filling fraction \eqref{ff1}. 
This is not always the case. However, it can be demonstrated that 
$T^{(k)}(\beta^{(k)} ; N_k)$ does not need 
any explicit integration of the filling fraction. 
This is done by fully exploiting  the flow equation. 

The starting point is to observe that  the flow equation \eqref{Hkflow}
must satisfy the self-consistency condition
\be
\frac{\partial^2 \tilde{H}^{(k)}}{\partial \kappa_a\partial \kappa_{b}}
=\frac{\partial^2 \tilde{H}^{(k)}}{\partial \kappa_{b} \partial \kappa_a}
~~~(0 \le a, \, b\le k-2) \,.
\label{consistency}
\ee 
This trivially looking conditions provide a very powerful tool to 
find $H_n ^{(k)}$. For example, the self-consistency shows that 
$H_0^{(k)}$ is a constant and is  independent of any $\kappa_\ell$.  
Not only that, the consistency condition 
turns out to constrain all the expectation values $\langle  \sum_i 1/\xi_i^t \rangle$
needed for  $ H_{-1}^{(k)}$.  

Let us redefine the expectation value  using a new  parameter 
\be 
\td_s    \equiv  - 2 \hbar b \beta^{(k)} \sum_{t=s+1}^{k}{\kappa_t} 
\left\langle  \sum_{i=1}^{N_k}  \frac1{\xi_i^{t-s}}   \right\rangle 
\ee
and put the flow equation in terms of $\td_s$ 
\begin{align}
\frac{\partial \tilde{H}^{(k)}}{\partial \tau_k}
&=-\frac{1}{\tau_k^2} \sum_{s=0}^{k-1} \td_{s} 
\left(  \sum_{j=k-s}^{k}
\frac{\kappa_j}{j} A_{j-(k-s)} \right) 
\label{ataum}\\ 
\frac{\partial \tilde{H}^{(k)}}{\partial \kappa_t}
&=\frac{1}{t \tau_k} \sum_{s=k-t}^{k-1} \td_{s} \, A_{s-(k-t)} 
 \label{akappa}
\end{align}
where 
$A_\ell=-(A_{\ell-1}+\kappa_{k-2} A_{\ell-2}+\cdots+\kappa_{k-\ell} A_{0})$
and $A_{0}=1$.

To make the consistency condition more tractable, 
we expand $\td_s$  in power series of $\{ \kappa_\ell \}$
since  $\vev{1/\xi_i^t} =  \CO(\{\kappa_k ^0 \})$.
\be
\td_s  =\sum_{\{\alpha_\ell \geq0 \}}
[\td_s]_{\alpha_0, \alpha_{1}, \cdots, \alpha_{k-2}} 
\tau_k^{\alpha_0} \kappa_1^{\alpha_1}  \cdots \kappa_{k-2}^{\alpha_{k-2}} \,.
\ee
The power expansion allows one to identify $H_{\ell}^{(k)}$ explicitly. 
Comparing \eqref{Hkflow} with 
\eqref{ataum} and \eqref{akappa} one has $H_{-1}^{(k)}$ 
using the zeroth order of $\tau_k$ in $\td_s$.
\be
H_{-1}^{(k)}=\sum_{s=0}^{k-1} \sum_{j=k-s}^{k} \sum_{\{\alpha_\ell \geq0 \}}
\frac{\kappa_j}{j} A_{j-(k-s)} 
[\td_s]_{0, \alpha_{1}, \cdots, \alpha_{k-2}} 
\kappa_1^{\alpha_1}  \cdots \kappa_{k-2}^{\alpha_{k-2}} \,.
\ee

Let us find out how useful the flow equation is. 
When $k=2$, we have only one flow equation 
with  $\tau_2=\eta_1$. 
\be
\frac{\partial \tilde{H}^{(2)}}{\partial \tau_2}
=-\frac{\td_0+\td_1}{2 \tau_2^2}\,.
\ee 
Once $\tau_2$ dependence of $(\td_0+\td_1)$ is known, 
one find $ \tilde{H}^{(2)}$ completely. 
For this, the loop equation provides a useful information.
The loop equation for $T^{(k)}$ is given as. 
\be
\tilde{f}(z) = 4\tilde{W}(z)^2-4\tilde{V}'(z) \tilde{W}(z) +2 \hbar Q \tilde W'(z) -\hbar^2 
\tilde W(z,z)
\label{tloop}
\ee
where $2 \tilde{W}(z)={\hbar b}  \vev{\sum_{i=1}^{N_k}1/(z-\xi_i) }$
and $\tilde W(z, w) = -b^2 \vev{\sum_{i,j=1}^{N_k}\frac1{(z-\xi_i)(w-\xi_j)} }_c$
is the connected 2-point resolvent. 
Then the quantum correction  $\tilde{f}(z)$ contains $\td_s$'s 
\be
\tilde{f}(z)=2 \hbar b \left\langle \sum_{i=1}^{N_k} 
\frac{-\tilde{V}'(z)+\tilde{V}'(\xi_i)}{z-\xi_i} \right\rangle = 
\frac{1}{\tau_k} \sum_{s=0}^{k-1} \frac{\td_s}{z^{2+s}} \,.
\ee 
One can collect the data of $\td_s$ after the large $z$ expansion 
of \eqref{tloop}  
\begin{align}
&[\td_0]_0 =0 \,~~~~[\td_0]_1 = \hbar bN_2 ( \hbar b N_2 + 2 \beta^{(2)}-\hbar Q )
\nn\\
&[\td_1]_0 =2\hbar b N_2 \beta^{(2)}\,,~~~[\td_1]_1 =-2 \hbar bN_2 ( \hbar b N_2 +  \beta^{(2)} -\hbar Q)\,. \nn
 \end{align} 
This simple data is enough to obtain the singular part 
\be
\tilde{H}^{(2)}=\frac{\hbar b N_2 \beta^{(2)}}{\tau_2}+
\frac{\hbar b N_2}{2}(\hbar b N_2-\hbar Q) \log \tau_2+\CO(\tau_2) 
\ee
and thus, we have 
\be
T^{(2)}(\beta^{(2)} ; N_2)=
\left( \frac{c_2}{c_1} \right)^{-b N_2( b N_2+2 {\beta^{(2)}}/{\hbar}-Q)}
\left( \tau_2 \right)^{-\frac{ b N_2 (b N_2-Q)}{2}}
e^{-\frac{ b N_2 \beta^{(2)}}{\tau_2 \hbar}+\CO(\tau_2)} \,.
\ee
If one puts $\beta^{(2)}=c_0$ and $Q=0$,
we reproduce the  result  in \eqref{sec3t2}.
This example shows that one may obtain 
the singular contribution of $T^{(k)}$ directly. 
We do not need  any explicit integration of the filling fraction as in 
section \ref{sec:solve} 
to connect the partition function with the filling fraction.
Here,  the flow equations together with the simple 
analytic structure of the loop equation is enough 
to find all the singular structures of the inner product. 

When $k=3$, there are two flow equations. Using the parameters $\tau_3=\eta_1 \eta_2^2$ and $\kappa_1=\eta_2$, we have 
the equations
\begin{align}
\frac{\partial \tilde{H}^{(3)}}{\partial \tau_3}
&=-\frac{(4 \kappa_1-1) \td_2+\td_1+2 \td_0}
{6 \tau_3^2} \label{3tau}\\
\frac{\partial \tilde{H}^{(3)}}{\partial \kappa_1} 
&= \frac{\td_2^{(3)}}{\tau_3}\,.
\end{align}
The  self-consistency ${\partial^2 \tilde{H}^{(3)}}/{\partial \tau_3 \partial \kappa_1}
= {\partial^2 \tilde{H}^{(3)}}/{\partial \kappa_1 \partial \tau_3}$ 
 gives the recursion relation:
\be
(m+1)(  [\td_2]_{n,m+1}-  [\td_1]_{n,m+1}) =2 (3n+2m-1) [\td_2]_{n,m} \,.
\label{d-recursion}
\ee

To solve the recursion relations we need the initial conditions  
such as $[\td_1]_{n,\alpha_1}$ for $\alpha_1 \ge 1$
and $[\td_2]_{n,0}$ .  The initial conditions are provided by the loop equation \eqref{tloop}.
Note that $\td_s$ is given for arbitrary $k$
\begin{align}
\td_s&=\hbar^2 b^2 \tau_k  \sum_{t=0}^{s}
\biggl\langle \sum_{i=1}^{N_k} \xi_i^{~t} \biggr\rangle 
\biggl\langle \sum_{i=1}^{N_k} \xi_i^{~s-t}\biggr\rangle
+2 \hbar b \beta^{(k)} \sum_{t=0}^{s} \kappa_t
\biggl\langle \sum_{i=1}^{N_k} \xi_i^{~s-t} \biggr\rangle \nn\\
& ~~-(s+1) \hbar^2 b \, Q\, \tau_k 
\biggl\langle \sum_{i=1}^{N_k} \xi_i^{~s} \biggr\rangle
+ \hbar ^2 b^2 \tau_k \sum_{t+m=s} 
\biggl\langle \sum_{i=1}^{N_k} \xi_i ^t
\sum_{j=1}^{N_k} \xi_j^m  \biggr\rangle 
 \,.
\label{trelation}
\end{align}
Using the fact $ \vev{\sum_{i=1}^{N_k}\xi_i} =-N_k + \CO(\kappa_\ell)$, 
we have at the zero-th order of $\tau_k$ 
\begin{align}
&[\td_s]_{0, \cdots ,0,\alpha_t =1,0, \cdots ,0}=(-1)^{s-t} 
2 \hbar b N_k \beta^{(k)}\quad (1 \leq t \leq s \le k-2 )
 \label{ds2}\\
&[\td_{k-1}]_{0, \cdots ,0}=2 \hbar b N_k \beta^{(k)}
 \label{ds4} \\
&[\td_{k-1}]_{0, \cdots ,0,\alpha_t =1,0, \cdots ,0}=(-1)^{k-1-t} 
2 \hbar b N_k \beta^{(k)}\quad (1 \leq t \leq k-2) \,.
\label{ds5} 
\end{align}
In addition, unless $\alpha_s=1$ and $\alpha_{s+1}=\cdots=\alpha_{k-2}=0 $ we have 
\be [\td_s]_{0, \cdots ,0, \alpha_s, \alpha_{s+1}, \cdots, \alpha_{k-2}} \equiv 0
~~~{\rm for~}  s \leq k-2\,.
\label{ds3}
\ee
At the first order of $\tau_k$, we have
\be 
[\td_s]_{1, 0, \cdots ,0}=(-1)^s \hbar b N_k
\left( (s+1) \hbar b N_k+2 \beta^{(k)} -(s+1) \hbar \,Q  \right) 
\label{ds1} \,.
\ee

Since $H_0^{(k)}$ is a constant, independent of $\kappa_\ell$'s, 
one can find $H_0^{(k)}$ using the flow  equation \eqref{ataum}.
From the first order of $\tau_k$ in $\td_s$, one has
\be
H_{0}^{(k)}=-\sum_{s=0}^{k-1} \sum_{j=k-s}^{k} \sum_{\{\alpha_\ell \geq0 \}}
\frac{\kappa_j}{j} A_{j-(k-s)} 
[\td_s]_{1, \alpha_{1}, \cdots, \alpha_{k-2}} 
\kappa_1^{\alpha_1}  \cdots \kappa_{k-2}^{\alpha_{k-2}} \,.
\ee
All the terms which depend on $\kappa_\ell$ should be cancelled.
Noting that $\kappa_{k-1}=\kappa_k=1$ and 
$A_\ell=(-1)^{\ell}+\CO(\{\kappa_\ell \})$
and changing the order of the sum over $s$ and $j$, one obtains
\be
H_0^{(k)}=-\sum_{j=0}^{1} \sum_{\ell=0}^{k-1-j} 
\frac{(-1)^{\ell} }{k-j} [\td_{j+\ell}]_{1, 0, \cdots, 0} 
=\frac{\hbar b N_k }{2} (\hbar b N_k - \hbar Q) \,. 
\label{apart}
\ee 

Now back to the case $k=3$,  we have  data from the loop equation
such as 
$[\td_1]_{0,m}=-2\hbar b N_3 \beta^{(3)} \delta_{m,1}$ and 
$[\td_2]_{0,0}= 2\hbar b N_3 \beta^{(3)}$. 
Solving the recursion relation \eqref{d-recursion},
one has 
$[\td_2]_{0,1}= -2\hbar b N_3 \beta^{(3)}$ 
which is consistent with the data \eqref{ds5} 
already  obtained from the  loop equation. 
Furthermore, $[\td_2]_{0,2}=-2\hbar b N_3 \beta^{(3)}$.
Finally the flow equation shows that 
\be
\tilde{H}^{(3)}=-\frac{\hbar b \beta^{(3)} N_3}{3 \tau_3}+2 \hbar b \beta^{(3)} N_3 
\frac{\kappa_1}{\tau_3}-\hbar b \beta^{(3)} N_3 \frac{\kappa_1^{~2}}{\tau_3}
+\frac{\hbar b N_3}{2} (\hbar b N_3-\hbar Q) \log \tau_3+\CO(\tau_3) \nn
\ee
and therefore, 
\be
T^{(3)}(\beta^{(3)};N_3)=
\left( \frac{c_3}{c_2} \right)^{-b N_3( b N_3+2 \frac{\beta^{(3)}}{\hbar}-Q)}
\left( \tau_3 \right)^{-\frac{ b N_3 (b N_3-Q)}{2}}
e^{-\frac{b \beta^{(3)} N_3}{\tau_3 \hbar} \left( 
-\kappa_1^{2}+2 \kappa_1-\frac{1}{3} \right)+\CO(\tau_3)} \,.
\ee
The singular structure is the same as the one in \eqref{sec3t3} 
where  $\beta^{(3)}=c_0$ and $Q=0$ are used.  

In general, one can prove that $T^{(k)}$ can be obtained 
from the consistency condition together with the loop equation for $k \geq 3$. 
The proof is presented in the appendix. 
To present  this idea more concretely, 
we explicitly calculate the case $k=4$. 
The flow equations are given by
\begin{align}
\frac{\partial \tilde{H}^{(4)}}{\partial \tau_4}
&=-\frac{1}{12 \tau_4^2}
\left[3 \td_0+\td_1+(3\kappa_2-1)\td_2+(-4\kappa_2+9\kappa_1+1)\td_3 \right] \\
\frac{\partial \tilde{H}^{(4)}}{\partial \kappa_1}
&=\frac{\td_3}{\tau_4} \,, \quad  \qquad
\frac{\partial \tilde{H}^{(4)}}{\partial \kappa_2}
=\frac{\tilde{D}_2-\td_3}{2 \tau_4} \,.
\end{align}
where $\kappa_1=\eta_2 \eta_3^{2}$, $\kappa_2=\eta_3$ and $\tau_4=\eta_1 \eta_2^2 \eta_3^3$.
Using the consistency of the flow equations
\be
\frac{\partial^2 \tilde{H}^{(4)}}{\partial \tau_4 \partial \kappa_2}
= \frac{\partial^2 \tilde{H}^{(4)}}{\partial \kappa_2 \partial \tau_4} \,,
\qquad \frac{\partial^2 \tilde{H}^{(4)}}{\partial \kappa_1 \partial \kappa_2}
= \frac{\partial^2 \tilde{H}^{(4)}}{\partial \kappa_2 \partial \kappa_1} \,,
\qquad \frac{\partial^2 \tilde{H}^{(4)}}{\partial \tau_4 \partial \kappa_1}
= \frac{\partial^2 \tilde{H}^{(4)}}{\partial \kappa_1 \partial \tau_4} \,,
\ee
we have the recursion relations:
\begin{gather}
2(m+1) [\td_3]_{n,\ell, m+1} = (\ell+1) [\td_2]_{n, \ell+1 ,m}-(\ell+1) [\td_3]_{n, \ell+1,m} \\
\begin{split}
(-12n-&9 \ell+3)[\td_3]_{n,\ell,m}+4(\ell+1) [\td_3]_{n,\ell+1,m-1}-(\ell+1) [\td_3]_{n, \ell+1,m} \\
&=(\ell+1) [\td_1]_{n,\ell+1,m}+3(\ell+1)[\td_2]_{n,\ell+1,m-1}
-(\ell+1)[\td_2]_{n, \ell+1,m}
\end{split} \\
\begin{split}
(6n+&4m-2) [\td_3]_{n,\ell,m}-(m+1) [\td_3]_{n,\ell,m+1}-9(m+1) [\td_3]_{n,\ell-1,m+1} \\
&=(m+1) [\td_1]_{n, \ell, m+1}-(m+1)[\td_2]_{n,\ell,m+1}
+(6n+3m-3)[\td_2]_{n,\ell,m}
\end{split}
\end{gather} 

The coefficients needed for  the singular part $H_{-1}^{(4)}$ 
are $[\td_s]_{0,0,m}$ with $m=0,1,2,3$ and $[\td_s]_{0,1,m}$ with $m=0,1$.
The initial data are given in \eqref{ds2} - \eqref{ds3};
$[\td_1]_{0,\alpha_1, \alpha_2}$,  $ [\td_2]_{0,0,\alpha_2} $ 
for arbitrary $\alpha_1, \alpha_2 \in \mathbb{Z}^+$
and $ [\td_2]_{0,1,0}$, $[\td_3]_{0,0,0} $, $ [\td_3]_{0,1,0}$, $ [\td_3]_{0,0,1}$.  
The recursion relation are solved to have 
$[\td_2]_{0,1,1}=-2 \hbar b N_4 \beta^{(4)}$, $[\td_3]_{0,1,1}=6 \hbar b N_4 \beta^{(4)}$,
$[\td_3]_{0,0,2}=-2 \hbar b N_4 \beta^{(4)}$, $[\td_3]_{0,0,3}=-4 \hbar b N_4 \beta^{(4)}$.
Therefore, 
\be
\tilde{H}^{(4)}=\frac{\hbar b N_4 \beta^{(4)}}{\tau_4} \left( 2 \kappa_1
-2\kappa_1 \kappa_2-\kappa_2+\kappa_2^2
+\frac{\kappa_2^3}{3}+\frac{1}{6} \right)
+\frac{\hbar b N_4}{2} (\hbar b N_4-\hbar Q) \log \tau_4+\CO(\tau_4) \nn
\ee
\be
\begin{split}
T^{(4)}(\beta^{(4)};N_4)=&\left( \frac{c_4}{c_3} \right)^{-b N_4( b N_4+2 {\beta^{(4)}}/{\hbar}-Q)}
\left( \tau_4 \right)^{-\frac{ b N_4 (b N_4-Q)}{2}} \\
& ~~~~~~~~~~~
e^{-\frac{b N_4 \beta^{(4)}}{\hbar \tau_4} \left( 2 \kappa_1
-2\kappa_1 \kappa_2-\kappa_2+\kappa_2^2
+\frac{\kappa_2^3}{3}+\frac{1}{6} \right)
+\CO(\tau_4)} \,.
\end{split}
\ee

Using the hierarchical structure \eqref{part_hier}, we obtain the 
singular part of the partition function
\be
\begin{split}
& \vev{\Delta|I^{(4)}(c_0)}_S
=\prod_{k=1}^{4} T^{(k)}(\beta^{(k)};N_k)   \\
&~~=(c_1)^{-h_1/\hbar^2+b N Q}
(\eta_1)^{-b (N_2+N_3+N_4) \left( b (N_2+N_3+N_4)+2 c_0/\hbar -Q \right)
-\frac{b^2}{2}  (N_2^2+N_3^2+N_4^2)+\frac{b Q}{2}(N_2+N_3+N_4)} \\
&~~~ \times
(\eta_2)^{-b (N_3+N_4) \left( b (N_3+N_4) +c_0/\hbar -Q \right)- b^2 (N_3^2+N_4^2)
+bQ(N_3+N_4)}
(\eta_3)^{-b N_4 \left( b N_4+ c_0/\hbar -Q \right)- \frac{3}{2}b N_4 (b N_4-Q)} \\
&~~~\times
 e^{-\frac{b c_0}{\hbar \eta_1}
\left( (N_2-N_3)+\frac{2 (N_3-N_4)}{\eta_2}-\frac{N_3-N_4}{3\eta_2^2}
+\frac{2 N_4}{\eta_2 \eta_3}+\frac{N_4}{\eta_2^2 \eta_3}
-\frac{N_4}{\eta_2^2 \eta_3^2}+\frac{N_4}{6 \eta_2^2 \eta_3^3} \right)} \,.
\end{split} \label{parti4}
\ee

\section{Summary and Discussion}\label{conclusion}
In this paper, the inner product of the irregular vector 
is studied which corresponds to the asymptotically free quivers 
of $SU(2)$ gauge groups (general theories of $A_1$ class) 
\cite{W1997, GMN2009, G2009N}.
The irregular vector  is the simultaneous 
eigenstate of a set of positive Virasoro generators. 
 The inner product of irregular vectors is obtained 
using the colliding limit of $(m+2)$-regular conformal block
and  is represented 
by the $\beta$-deformed Penner-type matrix model.
The partition function becomes the two point correlation
of irregular conformal block of rank $m$ and contains $(m+1)$-parameters. 
We have found explicitly the $m$-parametric dependence 
of the partition function. 

As shown in section \ref{sec:solve},  
we use the loop equation of the matrix model 
and find the parameter dependence of the inner product explicitly.   
In this process we need to evaluate the contour integral 
which is needed to eliminate the expectation values
in terms of filling fraction and parameters. 
However, the contour integral  in general gives elliptic function
and the inverting process 
is very cumbersome  to express the expectation values 
in terms of parameters including the filling fraction. 
 
On the other hand, it is noted in section \ref{sec:hier}
that as far as the singular structure is concerned 
one may use a simple and powerful method.  
The method uses the flow equation of the partition function.  
The idea is based on the observation that the singular structure 
of the inner product is hierarchical.  
One finds that the singular part of the 
inner product between irregular vectors of rank $n$ and $m$ is factorized into 
those of inner product between regular and irregular vectors \eqref{irr-factor}. 
Furthermore, the singular structure of the irregular vector  rank $m$ 
can be factorized into those of lower ranks as shown in \eqref{success}
and \eqref{part_hier}. 
Based on this hierarchical structure of the singularity, 
all the singular features are described 
by the effective partition function called $T^{(k)}$ \eqref{Tk}

The advantage of using $T^{(k)}$ 
is that the self-consistency  of the flow equations 
is enough to find all the singular structures of the partition function.
We do not need the contour integration corresponding to the filling fraction.
Why this method works lies in finding the initial condition for the flow equations.  
It is noted that two singular contributions, 
$H_{-1}^{(k)}$ and $H_{0}^{(k)}$ are responsible to $T^{(k)}$. 
The initial conditions needed for the singular part  
are trivially found  from the analytical properties 
of the loop equation of the matrix model
and are summarized in \eqref{ds2}$-$\eqref{ds1}
which hold for all orders of large $N$ expansion.
As the result,  the singular part are determined exactly.
This shows that there are $(m-1)$-types of instantons
for the irregular state of rank $m$
and the corresponding filling fraction 
becomes the instanton number. 
The filling fraction is fixed during the colliding process
but the instanton energy changes.
Note that the instanton energy which is  linear in the filling fraction in 
$H_{-1}^{(k)}$ and $H_{0}^{(k)}$ is found exactly to the all orders of 
the large $N$. 
The term quadratic in the filling fraction 
comes simply  from the Vandermonde determinant.
The chemical potential obtained this way is related with 
the $B$-cycle of the resolvent \cite{CE2006,MMS2010,MMM1010}.

On the other hand, the behavior of the
regular contribution  $H_n^{(k)}$ for $n \ge 1$ is quite different.
The initial condition is not  found 
from the simple analytic structure of the loop equation. 
The initial condition needs to be found from other methods
such as filling fraction integration.
This is the reason why we can use the filling fraction 
integration in section \ref{sec:solve} to find the regular contributions.

Depending on the way of colliding limit, there may arise many-point 
irregular conformal block
and  more  parameters appear.
The irregular $n$-point conformal block 
has the Penner potential with 
singularity  at $n$ points.  
One may see the same hierarchy of the singularity structure 
similar to 2-point  irregular conformal block. 

Note that $(m+1)$-parameters describe the irregular vector of rank $m$.
Among them,  $m$-parameter dependence
is easily described in terms of the $m$-flow equations. 
However, the remaining one parameter dependence is not simple to find. 
As seen in the irregular vector with rank 1,
the remaining parameter dependence can be obtained 
from the colliding limit of the 3-point function of the regular conformal 
block. 
When $\beta=1$,  the resulting partition function 
reduces to the original Penner model \cite{P1988} and describes 
the pseudo Euler characteristics and is useful 
to understand $c=1$ string theory \cite{DV1991, CDL1991}.
Therefore, the remaining parameter dependence 
should describe the  generalized descendants of the irregular vector.
The recent attempt to understand  the certain limit of  the regular conformal block
in terms of Painlev\'e equation \cite{GIL2013, LLNZ2013} 
will be useful to understand the remaining parameter dependence.

\subsection*{Acknowledgments} 
We thank A. Zamolodchikov and O. Lisovyy to draw attention to their
recent works on Painlev\'e equation.
This work is partially supported by the National Research Foundation of Korea (NRF) grant funded by the Korea government (MEST) 2005-0049409.

\appendix

\section{Self-consistency of the flow equation for $\tilde{T}^{(k)}$ }\label{app}

The flow equations  of $\tilde H^{(k)} =- \hbar^2 \log \tilde{T}^{(k)}$ is given 
in  \eqref{ataum} and \eqref{akappa}  
\begin{align}
\frac{\partial  \tilde {H}^{(k)}}{\partial \tau_k}
&=-\frac{1}{\tau_k^2} \sum_{s=0}^{k-1} \sum_{j=k-s}^{k}
\frac{\kappa_j}{j} A_{j-(k-s)} \td_{s} \,, 
\\ 
\frac{\partial \tilde{H}^{(k)}}{\partial \kappa_t}
&=\frac{1}{t \tau_k} \sum_{s=k-t}^{k-1} A_{s-(k-t)} \td_{s} \,, 
~~~~ {\rm for~} 1\le t \le k-2
 \label{app-h-flow}
\end{align}
where $A_{0}=1$ and 
$A_\ell=-(A_{\ell-1}+\kappa_{k-2} A_{\ell-2}+\cdots+\kappa_{k-\ell} A_{0})$.  
We use the special notations for some parameters 
$\tau_k=\kappa_0$ and $\kappa_{k-1}= \kappa_{k}=1$
and $\kappa_\ell =0 $ for $\ell <0$. 
Therefore,  $A_\ell=(-1)^{\ell}+\CO(\{\kappa_\ell \})$.  
$\td_s$ is expanded in power series of 
$\tau_k, \kappa_1, \cdots, \kappa_{k-2}$;
\be
\td_s = \sum [\td_s]_{\alpha_0, \alpha_1 \cdots, \alpha_{k-2} }
\tau_k^{\alpha_0} \kappa_1^{\alpha_1} \cdots \kappa_{k-2}^{\alpha_{k-2}}\,.
\label{app-d-power-series}
\ee

As  demonstrated in section \ref{sec:hier} up to $k=4$, 
the self-consistency  of the flow equations \eqref{app-h-flow} 
can fix all the singular parts of  $\tilde{T}^{(k)}$. 
In this appendix, we prove that the self-consistency 
of the flow equations can determine 
all the singular parts of  $\tilde{T}^{(k)}$  even for $k>4$. 
To prove this we use the self-consistency  condition \eqref{consistency}
\be
\frac{\partial^2 \tilde{H}^{(k)}}{\partial \kappa_a\partial \kappa_{b}}
=\frac{\partial^2 \tilde{H}^{(k)}}{\partial \kappa_{b} \partial \kappa_a}
~~~(0 \le a, \, b\le k-2) \,.
\label{app-consistency}
\ee 

The flow equation shows that $\tilde H^{(k)} $ has the form \eqref{formofH}
\be
\tilde{H}^{(k)}(\{\kappa_\ell\})=
\frac{H^{(k)}_{-1}}{\tau_k}+H_0^{(k)} \log \tau_k+
\sum_{n \ge 1} H^{(k)}_n \tau_k^n\,.
\ee
This shows that  the singular contribution to $T^{(k)}$ 
is due to the terms  $H_{-1}^{(k)}$ and $H_{0}^{(k)}$.  
According to the consistency condition \eqref{app-consistency}, 
$H_{0}^{(k)}$ is a constant independent of $\kappa_\ell$'s
whose value is found in \eqref{apart}.  
Therefore, we are going to concentrate on finding $H_{-1}^{(k)}$
which  is completely fixed if 
one knows the $\td_s$ at the limit  $\tau_k \to 0$,
{\it i.e.}, $  [\td_s]_{\alpha_0, \alpha_1 \cdots, \alpha_{k-2} }$ 
with $\alpha_0 =0$.

To find  $\td_s$ from the consistency condition \eqref{app-consistency} 
one needs some elementary information on $\td_s$.  
This is obtained from the loop equation.
The loop equation has the form \eqref{tloop}
\be
4\tilde{W}(z)^2-4\tilde{V}'(z) \tilde{W}(z) + 2\hbar Q \tilde{W}'(z)  - \hbar ^2 \tilde W(z,z) =\tilde{f}(z) 
= \frac 1{\tau_k} \sum_{s=0}^{k-1} \frac{\td_s}{z^{2+s}} \,.
\label{app-tloop}
\ee
Large $z$ expansion provides useful results on $\td_s$ 
as shown in \eqref{trelation}. 
Some  information  we need are listed as follows: 
\begin{align}
&[\td_{k-1}]_{\alpha_0 =0, \alpha_1=0,  \cdots, \alpha_{k-2}=0}  
= 2\hbar b N_k \beta^{(k)} 
\label{app-ads1}\\
&[\td_s]_{0, \cdots ,0,\alpha_t  =1,0, \cdots ,0} =(-1)^{s-t} 
2 \hbar b N_k \beta^{(k)}\quad (1 \leq t \leq s \le k-2 ) 
\label{app-ads2}
\end{align} 
and for  $s \le k-2 $ 
\be
[\td_s]_{0, \cdots ,0, \alpha_s, \alpha_{s+1}, \cdots, \alpha_{k-2}} \equiv 0
\quad \textrm{unless} ~~ \alpha_s=1
\,, ~~\alpha_{s+1}=\cdots=\alpha_{k-2}=0
\label{app-ads3} \,.
\ee
Note that the information holds to the all order of large $N$.
We present how to find $\td_s$ in the following steps.\\
\\
{\bf Step [1]: Find  $\td_{k-1}$ in power series in  $\kappa_{k-2}$.}\\
Use the consistency  flow along $\tau_k$ and $ \kappa_{k-2}$ 
\be
\frac{\partial^2 \tilde{H}^{(k)}}{\partial \kappa_{k-2} \partial \tau_k }
=\frac{\partial^2 \tilde{H}^{(k)}}{\partial \tau_k \partial \kappa_{k-2}}
\ee 
and find  $ \td_{k-1}$ in  power series in $\kappa_{k-2}$. 
This will determine 
$ [\td_{k-1}]_{\alpha_0=0, \alpha_1=0, 
 \cdots, \alpha_{k-3}=0, \alpha_{k-2} } $ 
with $\alpha_{k-2} \ge  0$.

 The consistency condition gives
\be
\sum_{s=0}^{k-1} \sum_{j=k-s}^{k}
\frac{\kappa_j}{j}
 \frac{\partial}{\partial \kappa_{k-2}}
\Bigl(  A_{j-(k-s)} \td_s \Bigr)=
 -\frac{\tau_k}{k-2} \frac{\partial}{\partial \tau_k}
\Biggl(\sum_{s=k-t}^{k-1}
 A_{s-2} \td_s \Biggr)
 \label{ak-2}
\ee
When $\tau_k\to 0$,  RHS of \eqref{ak-2} obviously vanishes. 
In addition, the initial condition \eqref{app-ads3} shows that 
$\td_s=0$ for $0 \le s \le k-3$ when 
$( \tau_k, \kappa_1, \cdots, \kappa_{k-3})  \to 0$, 
and therefore, the equation \eqref{ak-2} simplifies  to 
\be
 \left( \sum_{j=k-2}^{k} \frac{\kappa_j A_{j-1}}{j} \right) \,\td'_{k-1}
+   \left( \sum_{j=k-2}^{k} \frac{\kappa_j A'_{j-1}}{j} \right) \,  \td_{k-1}
= -\sum_{j=k-2}^{k} \frac{\kappa_j}{j} (A_{j-2} \td_{k-2})'
\label{app-ak-2a}
\ee 
where $X'$ denotes derivative with respect to $\kappa_{k-2}$.
The  initial condition \eqref{app-ads2} shows that 
$\td_{k-2}=2 \hbar b N_k \beta^{(k)} \kappa_{k-2}$.  
Therefore, \eqref{app-ak-2a} is the inhomogeneous first oder 
equation of $\td_{k-1}$, which gives the simple recursion relation 
for the power series in $\kappa_{k-2}$. 
This fixes $\td_{k-1}$ in power series in  $\kappa_{k-2}$
with the initial  condition 
$ [\td_{k-1}]_{ \kappa_{k-2} =0}  = 2\hbar b N_k \beta^{(k)} $ 
as given in   \eqref{app-ads1}. \\

{\bf Step [2]: Find  $ \td_{k-2}$ and $\td_{k-1}$ 
 in power series in  $\kappa_{k-3}$ and $\kappa_{k-2}$.}\\
Next step is to use the consistency  flow along $\tau_k$,  $ \kappa_{k-3}$ 
and  $ \kappa_{k-2}$. 
\be 
\frac{\partial^2 \tilde{H}^{(k)}}{\partial \tau_k \partial \kappa_{k-3}}
=\frac{\partial^2 \tilde{H}^{(k)}}{\partial \kappa_{k-3} \partial \tau_k } 
\,, ~~~~
\frac{\partial^2 \tilde{H}^{(k)}}{\partial \kappa_{k-2} \partial \kappa_{k-3}}
=\frac{\partial^2\tilde{H}^{(k)}}{\partial \kappa_{k-3} \partial \kappa_{k-2}} \,.
\ee
This additional two equations provide the recursion relations for 
$ \td_{k-1}$ and  $ \td_{k-2}$  in power series of  $\kappa_{k-3}$ and  $\kappa_{k-2}$
when  $\{ \tau_k , \kappa_1, \cdots , \kappa_{k-4}\} \to 0$. 
Using the result obtained from the step [1], the coefficients 
$ [\td_{k-1}]_{\alpha_0=0, \alpha_1=0,  \cdots, \alpha_{k-4}=0, \alpha_{k-3}>0 ,\alpha_{k-2 } \ge 0  } $
and 
$ [\td_{k-2}]_{\alpha_0=0, \alpha_1=0,  \cdots, \alpha_{k-4}=0, \alpha_{k-3}> 0 ,\alpha_{k-2} \ge 0 } $
are determined.  \\

{\bf Step [3]: Inductive proof of finding $ \td_{\ell}$'s  in power series in  $\kappa_{t}, \cdots \kappa_{k-1}$.}\\
Suppose the coefficients of $\td_\ell$ 
\be
[\td_\ell]_{0,\cdots,0,\alpha_{t+1},\alpha_{t+2},\cdots,\alpha_{k-2}}
~~~{\rm for~} (t+2 \leq \ell \leq k-1) 
\label{app-adv} 
\ee
is known for $(\alpha_{t+1}, \cdots, \alpha_{k-2}) \ge 0$. 
The case $t=(k-3)$ is done in the step [1], which is true. 
Therefore, our  proof can be done using the inductive way. 

Suppose the coefficient in  \eqref {app-adv}  is  known for  a certain $t < k-3$.  
Our claim is that the consistency conditions 
\be 
\frac{\partial^2 \tilde{H}^{(k)}}{\partial \kappa_{t} \partial \tau_k }
= \frac{\partial^2 \tilde{H}^{(k)}}{\partial \tau_k \partial \kappa_{t}} 
\,, ~~~~
\frac{\partial^2 \tilde{H}^{(k)}}{\partial \kappa_{t} \partial \kappa_{a}} 
=\frac{\partial^2 \tilde{H}^{(k)}}{\partial \kappa_{a} \partial \kappa_{t}}
~~{\rm for~} (t+1 \leq a \leq k-2)  \label{aself1} \,,
\ee 
will fix the next coefficient
 $ [\td_\ell]_{0,\cdots,0,\alpha_{t},\alpha_{t+1},\cdots,\alpha_{k-2}}$ 
for $ (t+1 \leq \ell \leq k-1) $. 

To prove this claim, let us use the consistency conditions 
\eqref{aself1} to get
\begin{align}
\sum_{s=0}^{k-1} \sum_{j=k-s}^{k}
\frac{\kappa_j}{j}
 \frac{\partial}{\partial \kappa_t}
\Bigl(  A_{j-(k-s)} \td_s \Bigr)&=
 -\frac{\tau_k}{t} \frac{\partial}{\partial \tau_k}
\Biggl(\sum_{s=k-t}^{k-1}
 A_{s-(k-t)} \td_s \Biggr)
 \label{app-atauder}\\
\frac{\partial}{\partial \kappa_t}
\Biggl(\sum_{s=k-a}^{k-1}
 A_{s-(k-a)} \td_s \Biggr) &=
  \frac{a}{t}\frac{\partial}{\partial \kappa_a}
\Biggl(\sum_{s=k-t}^{k-1}
 A_{s-(k-t)} \td_s \Biggr) \,.
\label{app-akader}
\end{align}
We further reduce the above equations using the known information. 
Note that we are trying to find the solution at $\tau_k=0$ and $\{\kappa_1, \cdots , \kappa_{t-1}\} = 0$.
Therefore, we put  $\tau_k \to 0$ and discard RHS of \eqref{app-atauder}.
In addition, $\td_s = 0$ for $s=0, \cdots, t-1$
by the conditions  \eqref{app-ads2} and \eqref{app-ads3}.
Therefore, the non-vanishing components in the above equations 
are simplified. 
\begin{align}
\sum_{s=t}^{k-1} \sum_{j=k-s \ge t }^{k}
\frac{\kappa_j}{j}
 \frac{\partial}{\partial \kappa_t}
\Bigl(  A_{j-(k-s)} \td_s \Bigr) &=0 
 \label{app-atauder-1} \\
\frac{\partial}{\partial \kappa_t}
\Biggl(\sum_{s=t }^{k-1}
 A_{s-(k-a)} \td_s \Biggr) &=
  \frac{a}{t}\frac{\partial}{\partial \kappa_a}
\Biggl(\sum_{s=t}^{k-1}
 A_{s-(k-t)} \td_s \Biggr) \,. \label{app-akader-1}
\end{align} 
Note that  we put the lower limit of the 
summation $s=t$ in \eqref{app-akader-1}
using $A_\ell =0$ when $\ell <0$,
without  which  the lower limit  should be
$s={\rm max}(k-a, t)$ (LHS)
and  $s={\rm max}(k-t, t)$ (RHS). 

Note that  $\td_t$  is known and  linear in $\kappa_t$
in \eqref{app-ads2} and \eqref{app-ads3}.
The unknowns are $\td_s$ with $ (t+1) \le s \le (k-1)$. 
Therefore, it is convenient to put  the equations 
\eqref{app-atauder-1} and \eqref{app-akader-1}  
in a simple product form 
\be
\sum_{s=t+1}^{k-1} \tilde{A}_s \td_s'=B_t  \,,
~~~~~
\sum_{s=t+1}^{k-1} A_{s-(k-a)} \td'_s=B_a ~~
{\rm for~} (t+1 \leq a \leq k-2)
\label{app-adb}
\ee
where $X'$ denotes derivative with respect to $\kappa_{t}$.
$\tilde{A}_s$ is the weighted sum of $A_s$, 
$\tilde{A}_s = \sum_{j=k-s}^{k} ({\kappa_j A_{j-(k-s)}}/{j})$ 
and $ B_t $ and $B_a$ contain  $\td_s$'s
but no derivatives of $\kappa_t$'s. 
\begin{align}
B_t &= -\sum_{s=t}^{k-1} \sum_{j=k-s}^{k}
\frac{\kappa_j A'_{j-(k-s)} \td_s}{j}
-\tilde{A}_t \td'_t  \,, \nn \\
B_a &= \frac{a}{t} \frac{\partial}{\partial \kappa_a}
\Biggl( \sum_{s=k-t}^{k-1} A_{s-(k-t)} \td_s \Biggr)
-   \delta_{t, k-a} \, \td'_t \,. \nn
\end{align}
where $\td_t'= 2 \hbar N_k \beta^{(k)}$.  

Putting the new  equations in a  matrix form 
$\mathbb{A}  \mathbb{D}' = \mathbb {B} $, 
we have  $(k-t-1) \times (k-t-1)$  invertible matrix  $\mathbb{A}$ 
\be
\mathbb{A} := \left( 
\begin{array}{cccc}
\tilde{A}_{t+1} & \tilde{A}_{t+2} & \cdots  \cdots & \tilde{A}_{k-1} \\
A_{t-1} & A_t & \cdots  \cdots & A_{k-3} \\
A_{t-2} & A_{t-1} & \cdots  \cdots & A_{k-4} \\
\vdots & \vdots &  &\vdots \\
\vdots & \vdots &  &\vdots \\
A_{t-(k-t-2)} & A_{t-(k-t-3)} & \cdots  \cdots & A_t
\end{array} \right)
\ee
and $k-t-1$ column vectors  $ \mathbb{D}$ and $  \mathbb {B} $ 
\be 
 \mathbb{D} =   \left( 
\begin{array}{c}
\td_{t+1} \\ \vdots \\ \td_{k-1}
\end{array} \right) \,,~~~~
\mathbb{B} 
  = \left( 
\begin{array}{c}
B_t \\ \vdots \\ B_{k-2}
\end{array} \right) \,.
\ee 

Inverting the matrix equation  we have $  \mathbb{D}' = \mathbb{A}^{-1}   \mathbb {B} $. 
When $\mathbb{D}$ is put in power series of $\kappa_t$, 
$\mathbb{D} =\sum_{a \ge 0} \mathbb{D}_a \kappa_t^a $, 
the inverted equation provides the  recursion relation of the  $\mathbb{D}_a$'s.
which is solved if the initial condition $\mathbb{D}_0$ is known. 
Note that $\mathbb{D}_0$ is the assumption of our claim \eqref{app-adv}. 
Therefore, the claim is proved.  

Note that the iteration goes from $t=(k-3)$ to $t=1$. 
The procedure determines all the contribution to $\td_\ell$ 
($ 2 \leq \ell \leq k-1$)  when $\tau_k = 0$.  
The remaining $\td_1$ is already known  completely in 
\eqref{app-ads2} and \eqref{app-ads3}  when $\tau_k = 0$.


\end{document}